\documentclass[10pt,twocolumn,letterpaper]{article}

\usepackage{wacv}
\usepackage{times}
\usepackage{epsfig}

\usepackage{ragged2e}

\usepackage[nocompress]{cite}
\usepackage[T1]{fontenc}

\usepackage{aecompl}

\usepackage{graphicx}

\usepackage[caption=false,font=normalsize,labelfont=sf,textfont=sf]{subfig}

\usepackage{array}
\usepackage{multirow}
\usepackage{booktabs}
\usepackage{tabularx}

\usepackage{xcolor,colortbl}
\usepackage{color}

\usepackage{amsmath}
\usepackage{bbm}
\usepackage{amsfonts}
\usepackage{mathtools}

\usepackage{amsthm}

\usepackage{amssymb}

\usepackage[ruled,vlined]{algorithm2e}
\usepackage{algpseudocode}

\usepackage{paralist}

\usepackage{cuted}

\wacvfinalcopy %

\def\wacvPaperID{335} %

\begin{document}
	
	\title{IterNet: Retinal Image Segmentation Utilizing Structural Redundancy \\in Vessel Networks}

	\author{Liangzhi Li \\
		Osaka University\\
		{\tt\small li@ids.osaka-u.ac.jp}
		\and
		Manisha Verma \\
		Osaka University\\
		{\tt\small mverma@ids.osaka-u.ac.jp}
		\and
		Yuta Nakashima \\
		Osaka University\\
		{\tt\small n-yuta@ids.osaka-u.ac.jp}
		\and
		Hajime Nagahara \\
		Osaka University\\
		{\tt\small nagahara@ids.osaka-u.ac.jp}
		\and
		Ryo Kawasaki \\
		Osaka University\\
		{\tt\small ryo.kawasaki@ophthal.med.osaka-u.ac.jp}
	}
	
\makeatletter
\let\@oldmaketitle\@maketitle%
\renewcommand{\@maketitle}{
	
	\newpage
   \null
   \vskip .375in
   \begin{center}
      {\Large \bf \@title \par}
      \vspace*{15pt}
      {
      \large
      \lineskip .5em
      \begin{tabular}[t]{c}
         \ifwacvfinal\@author\else Anonymous WACV submission\\
         \vspace*{1pt}\\
Paper ID \wacvPaperID \fi
      \end{tabular}
      \par
      }
      \vskip .5em
      \vspace*{12pt}
   \end{center}
   
	\vspace{-5mm}
	\centering
	\includegraphics[width=0.9\textwidth]{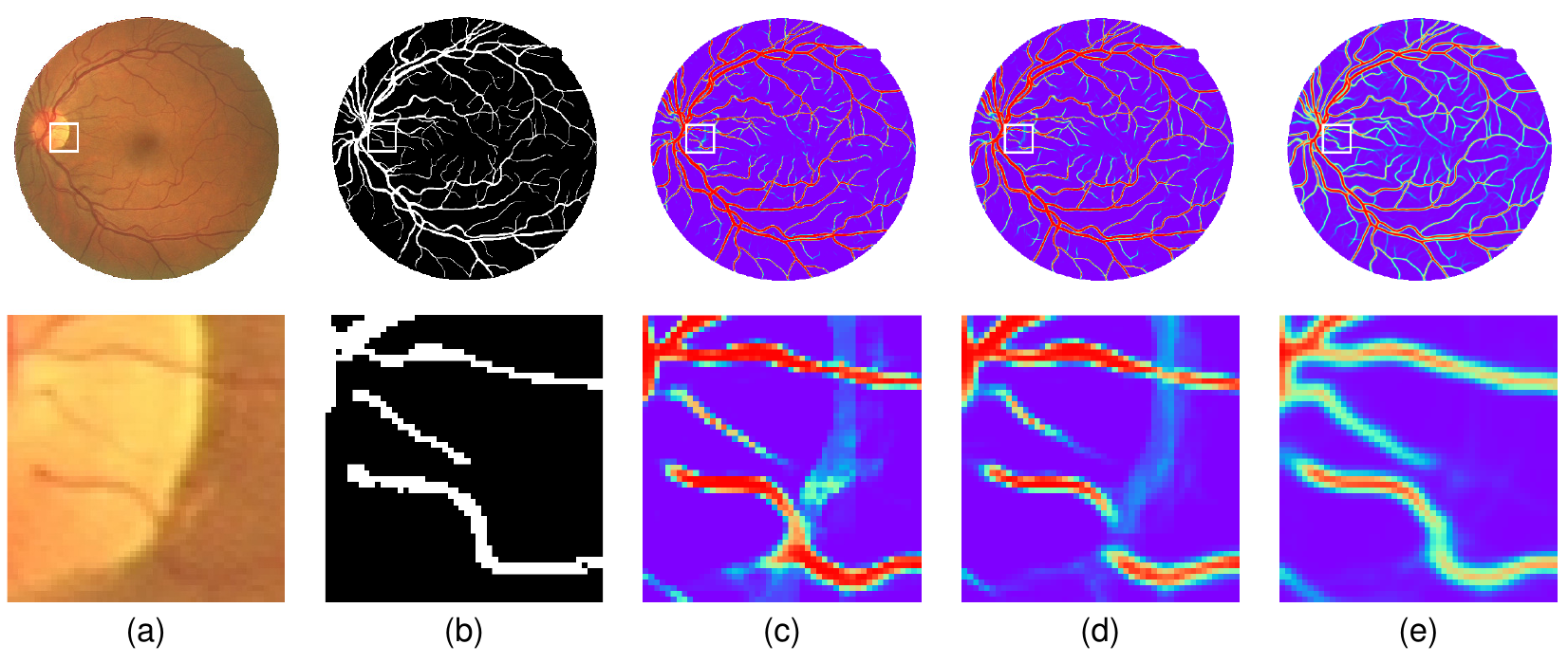}
	
	\justify
	\refstepcounter{figure}\normalfont Figure~\thefigure. IterNet analyzes the vessel network in a retinal image for fine segmentation. The first row is the whole image and the second row is an enlarged image of an area near the bright spot. Red color means a high possibility for a pixel to be part of the vessel while blue color represents a low possibility. We can see that IterNet well handles incomplete details in the retinal image and infers the possible location of the vessels. (a) An example image from the DRIVE dataset, (b) The gold standard, (c) UNet (AUC: 0.9752), (d) Deform UNet (AUC: 0.9778) and (e) IterNet (AUC: 0.9816).
	\label{fig_story}
	\vspace{4mm}
}
\makeatother

	\maketitle
	\thispagestyle{plain}

\begin{abstract}
	\vspace{-3.5mm}
 Retinal vessel segmentation is of great interest for diagnosis of retinal vascular diseases. To further improve the performance of vessel segmentation, we propose IterNet, a new model based on UNet \cite{UNet}, with the ability to find obscured details of the vessel from the segmented vessel image itself, rather than the raw input image. IterNet consists of multiple iterations of a mini-UNet, which can be 4$\times$ deeper than the common UNet. IterNet also adopts the weight-sharing and skip-connection features to facilitate training; therefore, even with such a large architecture, IterNet can still learn from merely 10$\sim$20 labeled images, without pre-training or any prior knowledge. IterNet achieves AUCs of 0.9816, 0.9851, and 0.9881 on three mainstream datasets, namely DRIVE, CHASE-DB1, and STARE, respectively, which currently are the best scores in the literature.
The source code is available\footnotemark.
\vspace{-5mm}
\end{abstract}
\footnotetext{Source code: https://github.com/conscienceli/IterNet}

\section{Introduction}\label{section_intro}
Retinal examination serve as an important diagnostic modality in finding retinal diseases as well as systemic diseases, such as high blood pressure, arteriolosclerosis, and diabetic retinopathy, a microvascular complications of diabetes. In fact, it is the only feasible way for the doctors to inspect the blood vessel system in the human body \textit{in vivo}. It has been used as a routine examination not only by ophthalmologists but also many other specialists \cite{chatziralli2012value}.
Retinal examination is non-invasive and economical to perform, and it has been widely conducted all over the world. However, at the same time, there will be a huge gap between the needs and the capacity of handling the ever-increasing retinal images by ophthalmologists. Computer-aided diagnosis will be an obvious solution in this scenario, and vessel segmentation is the essential basis of following analysis.

\begin{figure*}
	\centering
	\includegraphics[width=\textwidth]{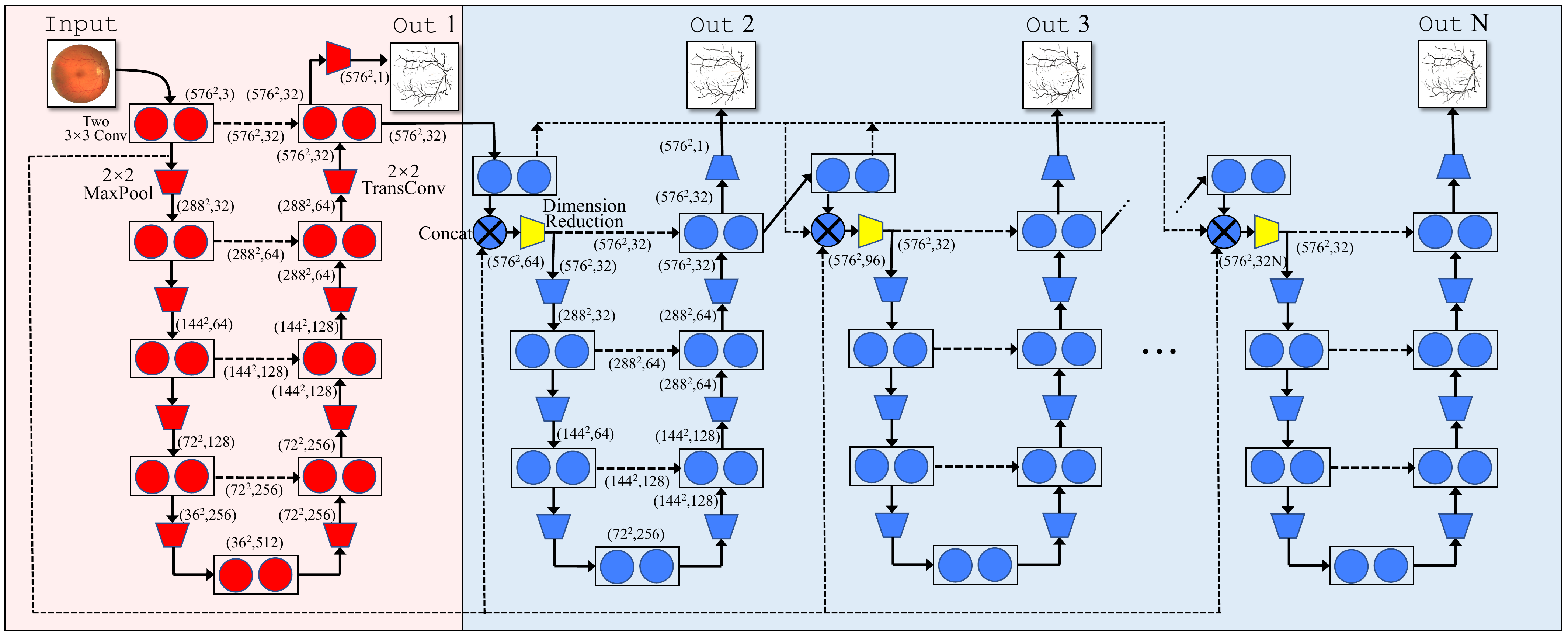}
	\caption{The structure of IterNet, which consists of one UNet and iteration of $N-1$ mini-UNets.}
	\label{fig_structure}
\end{figure*}

One major difficulty in the vessel segmentation task is that vessels have no significant differences in appearance from the background, especially for the micro vessels in noisy images. It is challenging to find every vessel while not introducing too many false positives. Actually, things will be more complicated if we consider the problem of photo imaging quality. Much essential information may be lost due to improper illumination, sensor noises, \etc. In this case, it is indeed impossible for segmentation models to find a complete yet accurate vessel network. For example, in Fig.~\ref{fig_story}(a), because of the optic disk (brighter spot) in the image, segmentation results suffer from severe deterioration around its boundary: Some pixels have been ``lost'' in the large gap in luminance. 

Figure \ref{fig_story}(b) is the gold standard marked by human experts, who can do this because they know vessels should be lines/curves, which should be connected to each other to form a network. In other words, this \emph{structural redundancy} enables the experts to interpolate vessels in obscured regions in retinal images. Deep learning models may also be capable of learning this kind of knowledge if they are exposed to a large amount of perfectly labeled data, which are extremely limited in the retinal image segmentation field. In fact, there are no more than 20 images for training in publicly available datasets, \ie, DRIVE \cite{staal:2004-855}, CHASE-DB1 \cite{owen2009measuring}, and STARE \cite{845178}. %

Existing approaches struggle with this scarceness of data. As shown in Figs.~\ref{fig_story}(c) and (d), the two state-of-the-art models with top performance, namely UNet \cite{UNet} and Deform UNet \cite{JIN2019}, face noticeable errors in their prediction. They either mix up the vessel and the boundary of optic disk or fail to detect vessels around their intersection, leading to \eg, segmentation in which a single vessel is split into two unconnected parts. This is a common phenomenon in medical image segmentation and can lead to a defective vessel map consisting of a set of disconnected or broken up segments. This issue makes it very difficult to analyze the blood vessel condition by doctors or standard imaging methodologies using the segmented images \cite{7319356}. Therefore, connectivity is also an important problem for retina segmentation.

One interesting observation in Figs.~\ref{fig_story}(c) and (d) is that humans may still be able to infer where the actual vessels are from these resulting vessel maps. This is because, like the experts, we can also make use of structural redundancy; we can guess that two parts in predicted vessels are connected if their edges are close and pointing to each other. This may also apply to deep learning models. Although it is hard for deep learning models to directly overcome the problem of missing or extra predictions, it may be possible to let them know which segmented vessel is false and which is not. Consequently, they may be able to learn how to fix errors in segmentation results. Based on this observation, we design a new UNet-based model, coined \textit{IterNet}, which can well utilize the structural redundancy in the vessel system. The resulting vessel map by IterNet is shown in Fig.~\ref{fig_story}(e), which gives precise segmentation of the vessels and almost avoid the interference around the optical disk.

The key idea is to shift the focus of the deep learning model from dealing with every pixel in raw input images to the whole vessel network system. More specifically, we build a model that refines imprecise vessel segmentation results to more precise ones, but not directly maps raw input images to precise segmentation results. In order to let the model learn sufficient knowledge of what real vessel networks and ones with failure in segmentation results look like, it is essential to provide them with enough training samples. However, again, there are no datasets available for this sake as mentioned above. 

One feasible way is to use the outputs of a certain segmentation model, which actually is vessel maps, like the ones in Figs.~\ref{fig_story}(c) and (d), as inputs to the model dedicated for refinement. We implement this by adding some refinery modules (mini-UNets) after a base module (UNet) for initial segmentation, as shown in Fig.~\ref{fig_structure}. The input of each refinery module is the output of the second last layer of its preceding module. Each module has an output of vessel segmentation, which has a respective loss function. In training, the base module will consistently adjust its parameters to improve its own output. Therefore, the first refinery module will get virtually different inputs, even with a fixed number of training samples. This applies to other refinery modules as well. In this process, the refinery modules can be exposed to a large number of false vessel patterns and thus can learn how to fix them because they are all bound to the correct labels. The number of refinery modules is a hyperparameter to be tuned according to the number of training samples, GPU capacities, and training time. The output from the last module, ``\texttt{Out} $N$'' in Fig.~\ref{fig_structure}, will be the actual output in prediction and all other outputs are only used for training. In addition, to avoid the overfitting problem and to improve the training efficiency, we design IterNet with the weight-sharing feature and a skip-connection structure.%

The main contributions of our work are as follows.
\begin{compactitem}
	\item A vessel segmentation model with top performance over all mainstream datasets.
	\item An iterative design of neural network architecture to learn the nature of vessels, with avoiding overfitting by weight-sharing.
	\item Drastically improved connectivity of segmentation results.
\end{compactitem}

\section{Related Work}\label{Related_Works_section}

\paragraph{Image segmentation:} 
Currently, most state-of-the-art models \cite{DOLZ2018456,Zhang_2018_CVPR,Jegou_2017_CVPR_Workshops} for semantic segmentation stem from a fully-convolutional design, which is first introduced by the fully convolutional network (FCN) \cite{Long_2015_CVPR}.
The main idea is to encode the raw images into a feature space and convert feature vectors into segmented images in an end-to-end manner. %
FCN has innovated many iterative segmentation approaches, which have similar ideas with our IterNet, but with totally different implementations. For example, the iterated interactive model \cite{Amrehn:2017:UIA:3309883.3309905} runs the FCN model several times and takes the users' feedback to add more accurate training labels during each iteration. Drozdzal's model uses FCN to preprocess input images into a normalized version and then applies a fully convolutional ResNet for iteratively refining the segmented images \cite{DROZDZAL20181}.
UNet \cite{UNet} is another well-known fully-convolutional model%
. Unlike FCN, UNet has multiple decoding layers to upsample the features. It also adds some skip-connections to allow decoding layers to use the features from the encoding process.%

\paragraph{Retinal image segmentation:}

The traditional way to conduct blood vessel segmentation is to utilize the local information, such as image intensity, or some hand-crafted features to perform classification. One earliest attempt is to use thresholding and masking. Roychowdhury \etal \cite{7042289} introduced an iterative segmentation method. Several processes in the segmentation algorithm run multiple times, which is very similar to our IterNet. Their method literately looks for the possible vessel pixels by adaptive thresholding on a retinal image, which is masked with the segmentation result obtained from the last iteration. %

The emergence of UNet \cite{UNet} leads to a new era of image segmentation in the medical domain, and has revolutionized most image segmentation tasks in relevant domain \cite{8341481,8681706,8589312,DBLP:journals/corr/abs-1903-00923,TANG2019289}. %
Kim \etal \cite{8036917} adopted the concept of iterative learning in an UNet-like model. Being similar to IterNet, their model also uses the last output as the next input. The main difference from ours is that they simply run one same model for multiple times. The encoding and decoding modules still need to deal with both raw retinal images and vessel segmentation results. In contrast, IterNet is one single model with iterated mini-UNets, which completely separates raw image input and segmentation result input. This is the key design concept of IterNet, boosting the state-of-the-art performance. %
Another recent model \cite{8697107, 8379359}, named DenseBlock-UNet transforms the convolutional modules in the common UNet model into the dense block introduced in \cite{8099726}. The dense block can improve UNet in some aspects, like alleviating the gradient vanishing, strong feature propagation, enabling feature reuse, and decreasing the whole parameter size. %
Deform-UNet \cite{JIN2019} is another encouraging model. The authors modified the UNet model for better performance. They applied two key modules from the deform convolutional networks \cite{dai17dcn}, namely deformable convolution and deformable RoI pooling, which replace the original modules in standard convolutional neural network (CNN) models and empower them with the ability to dynamically adjust their receptive fields according to the actual objects in input images. %

One of the main differences between IterNet and other UNet-based models is that our focus is not on modifying the structure of UNet; we think the feature extraction ability of UNet is enough for the vessel segmentation task. %
We are instead trying to make a better use of well-extracted features from the UNet model to infer missing pieces in them.

\section{IterNet}\label{section_method}

Based on the observation mentioned in Section \ref{section_intro}, we design our model to learn what the human blood vessel system in retinal images looks like to exploit its structural redundancy. The network is designed by keeping human annotators in mind. That is, an annotator may segment a raw retinal image in  several stages: The first stage is to make a rough segmentation map. In the following stages, they keep improving the map with the help of raw retinal image and previous vessel map until the annotator is satisfied with the resulting vessel map.
This leads to the idea of using resulting vessel maps (as in Fig.~\ref{fig_story}(c)) from a base module as an input to a refinery module that learns to correct it. With this architecture, the refinery module can infer missing/extra predictions based on the structure of the vessel system. In order to complete correction, we can apply the refinery module iteratively as shown in Fig.~\ref{fig_structure}. 

More specifically, our network consists of two slightly different architectures: One is UNet, and the other is a simplified version of UNet, referred to as mini-UNet. We use UNet as our base module because of its superior performance in various segmentation tasks, especially in the medical applications. The output of UNet is the one-channel map of the probabilities of pixels being on a vessels. The refinery modules' architecture is mini-UNet, and they use the output of the second last layer of its precedent module, which is a 32-channel feature map and thus can have more information, compared with the one-channel vessel probability map. The mini-UNet actually is a light-weight version of the UNet architecture with fewer parameters because the input to the refinery modules is a feature map that we consider is simpler than the raw retinal images with all the background and noises. 
In addition, we conduct an experiment to test the performance when replacing mini-UNets with full-size UNets, and the results get worse on all three datasets (Refer to the supplementary material for detailed results).

\begin{figure}[!t]
	\setlength{\fboxsep}{0pt}%
	\setlength{\fboxrule}{0.2pt}%
	
	\centering
	\subfloat[]{\includegraphics[width=0.26\columnwidth, height=0.5\columnwidth]{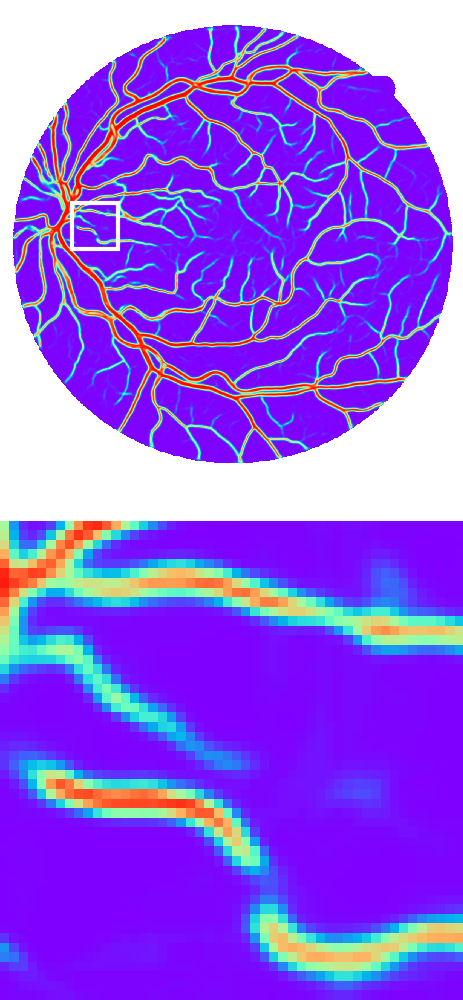}}
	\hfil
	\subfloat[]{\includegraphics[width=0.26\columnwidth, height=0.5\columnwidth]{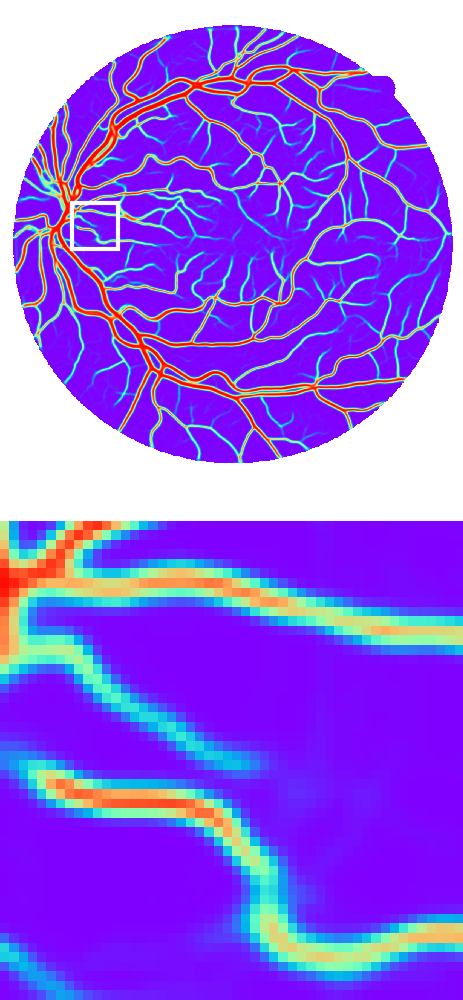}}
	\hfil
	\subfloat[]{\includegraphics[width=0.26\columnwidth, height=0.5\columnwidth]{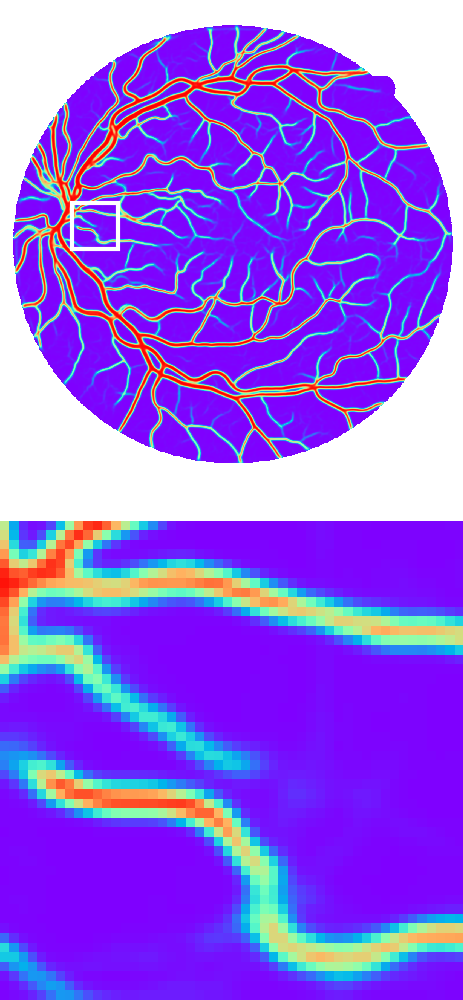}}
	\hfil
	\caption{The result of \texttt{Out} 1, 2, and 3 from IterNet. The corresponding AUCs are 0.9793, 0.9812, and 0.9815, respectively. }
	\label{fig_iter}
\end{figure}

As we can see in Fig.~\ref{fig_story}(c), the mapping from original retinal images to vessel maps is mostly learnt by the base module, and the refinery modules are responsible only for small parts of the vessels (\eg, thin vessels). Hence, IterNet achieves good segmentation results if we have enough samples to train the refinery modules. %
In our architecture, all refinery modules (the modules marked in blue in Fig.~\ref{fig_structure}) share the same weights and biases. The input of first refinery module is the feature map from the second last layer of the base module, and the rest refinery modules follow the similar procedure. Essentially, this can be interpreted as the same module running for multiple times in a single forward path. The most obvious benefit is that they can have varying inputs. The intermediate results of the vessel map always changes after each refinery module as illustrated in Fig.~\ref{fig_iter}. As a result, the refinery modules are consistently exposed to new patterns of failure in the vessel segmentation. This architecture makes it possible to train the refinery modules with only 20 training samples.

Another reason for this architecture is to use iterative prediction, which can improve the segmentation performance. We observe that one single model prefers to modify the results only by small differences, and the concept of iterated prediction has been used in many existing methods \cite{7042289,8036917}. Even with the same segmentation model, iterative applications can still get better results. Therefore, we can say that iterative application of the same model allows to refine the missing parts of the vessel network without explicitly modeling its structural redundancy. 
In Fig.~\ref{fig_iter}, we show the output from the base module and three outputs from following three refinery modules. The fourth output is shown in Fig.~\ref{fig_story}(e). We can see that, with the iterative predictions by the refinery modules, IterNet gradually connects split micro vessels together.

One important issue is that our IterNet is a many-layered feed-forward network. In general, upper-layers of a many-layered network hardly have an access to the input (or the features from layers close to the input layers), whereas it can serve as an important reference for the mini-UNets to see what the original vessels look like and make decisions based on it. Even for human annotators, it is necessary to check the specific area in the raw vessel images when refining some extremely-fuzzy parts of the vessel network. Therefore, we should enable the higher-layers to utilize the features from the lower layers. In addition, deep learning models may suffer from the vanishing gradient problem when they are deep. Hence IterNet demands paths from the upper layers to lower layers for efficient back-propagation. 

We therefore add some skip-connections to IterNet, similar to common UNet. There are three kinds of skip-connections in IterNet. The first one is the intra-module connection to connect the encoding layers of each to the decoding layers. The second one is from the base UNet to all refinery mini-UNets. This connection provides an access to the feature from the first layer of the base UNet, which is very close to the input retinal image. The feature is concatenated with the feature from the first layer of every mini-UNet. The third one is the connections among the mini-UNets, inspired by the dense connection of the dense network \cite{8099726}. The features from lower modules are concatenated with those from the upper modules. To keep the same structure and for weights-sharing among the mini-UNets, we add a $1\times 1$ convolutional layer, which is marked in yellow in Fig.~\ref{fig_structure}, for dimensionality reduction. This is the only component in the mini-UNets that has private parameters.

For training IterNet, we employ losses for each output \texttt{Out} $i$. We use the sigmoid cross entropy, defined as:
\begin{equation}
    L_i = -y_i \log(p_i) - (1-y_i)\log(1-p_i),
\end{equation}
where $y_i$ represents the binary indicator (0 or 1) whether the label is correct for the pixel $i$, and $p_i$ is the predicted probability that the pixel $i$ is a vessel pixel. Then they are summed up with certain weights as:
\begin{equation}
    L = \sum_i w_i L_i
\end{equation}
where $w_i$'s are set to 1 as we put no particular importance to any output.

\section{Implementation Details} \label{section_details}

\subsection{Data Augmentation} 

As the number of training images is no more than 20 in publicly-available common datasets, some augmentation techniques are necessary to avoid overfitting. We attempt to feed the IterNet model with all possible variations, including color, shape, brightness, and position, to make the model adapt to various imaging sensors, environments, color ranges, \etc. We use a training sample generator to consistently produce randomly modified samples during the training process. 

\subsection{Image patches in Training and Prediction}

It is common to divide an input image into image patches in the same size on a regular grid, which increase the number of available training samples. As IterNet has no requirement on the consistency of the input image sizes. There are three different ways for training and prediction: 
\begin{compactitem}
    \item Use image patches for training and testing, conquering the resulting image patches together as the final result. This strategy may make the best use of the training material and gave the most refined prediction results. However, it will cost much longer time than the other two methods because inference process has to be conducted for many times (see the supplementary material for detailed time cost).
    \item Use image patches only in training, and use the whole image in prediction to directly get the final result. This strategy can also use augmented training data. However, it may not perform as well as the first strategy in prediction because the model is trained with image patches.
    \item Use the original image in both training or prediction, which is seldom adopted because the available retina data are very limited; therefore, data augmentation usually helps.
\end{compactitem}
We employ image patch size of 128 pixels for training IterNet to avoid overfitting. For prediction, we test both whole image prediction and image patch prediction.

\begin{figure}[!t]
	\setlength{\fboxsep}{0pt}%
	\setlength{\fboxrule}{0.2pt}%
	
	\centering
	\subfloat[]{\includegraphics[height=0.5\columnwidth]{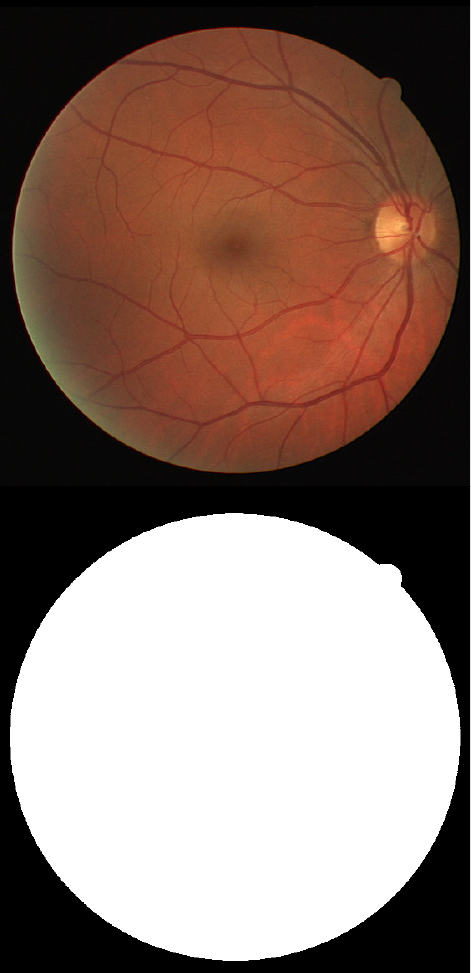}}
	\hfil
	\subfloat[]{\includegraphics[height=0.5\columnwidth]{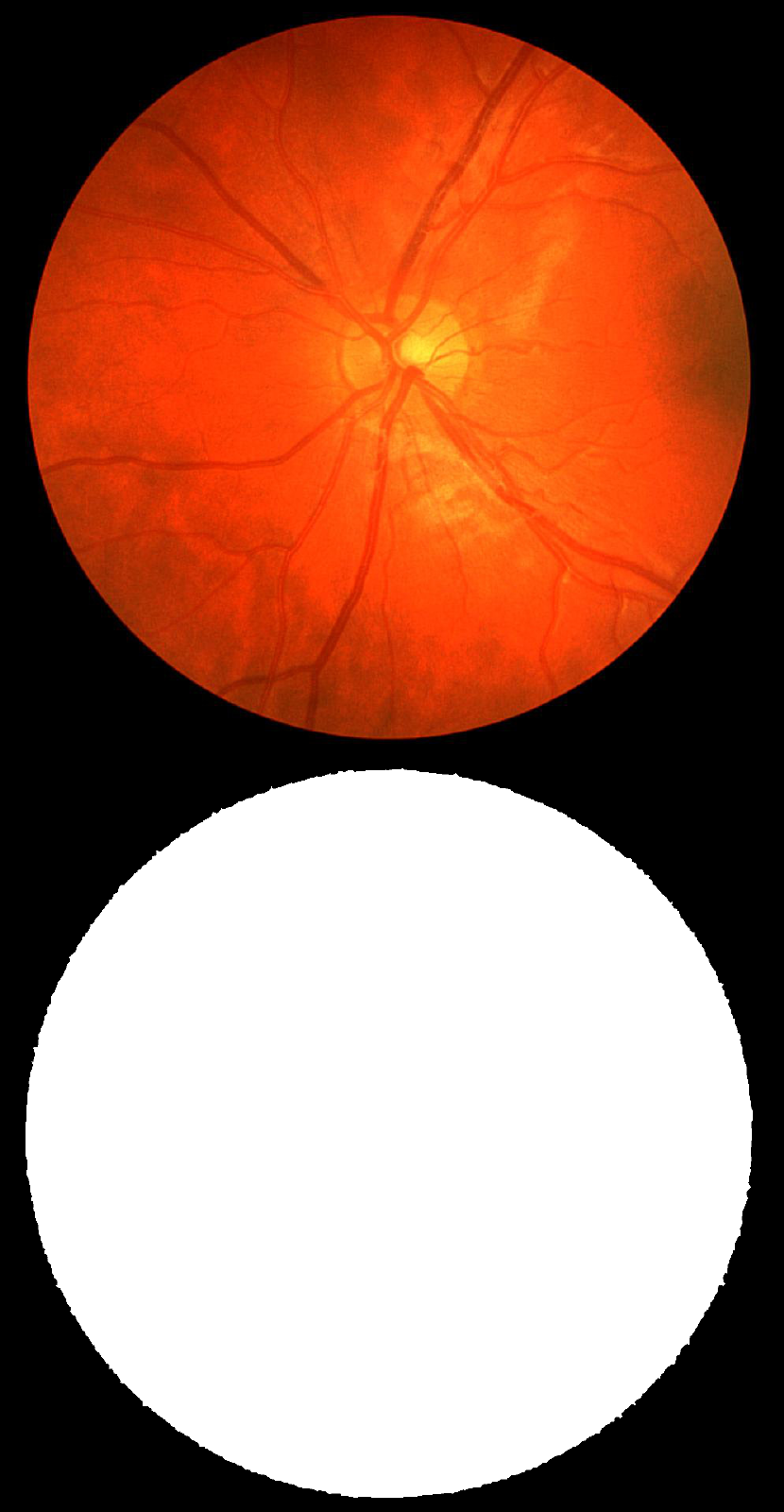}}
	\hfil
	\subfloat[]{\includegraphics[height=0.5\columnwidth]{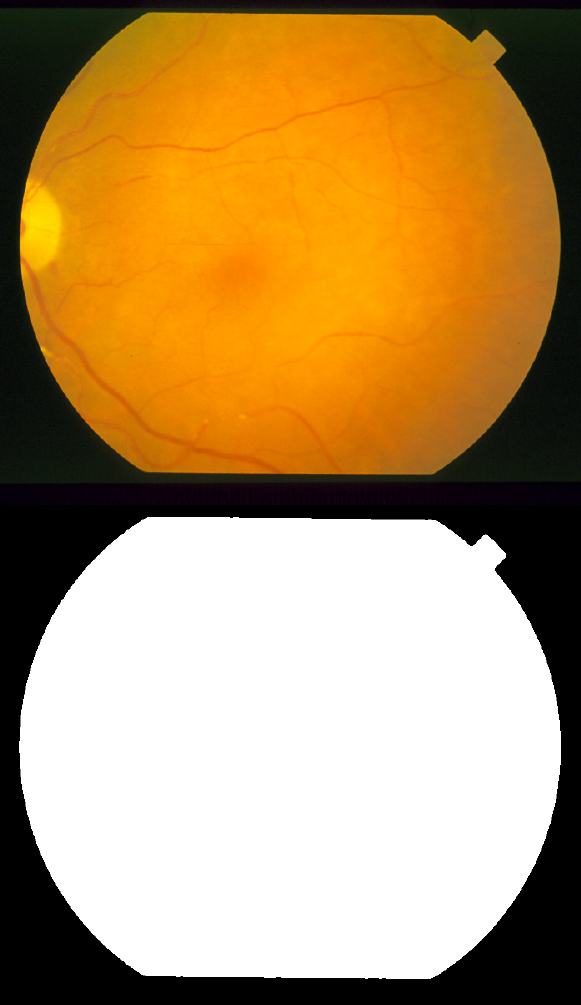}}
	\hfil
	\caption{Raw images and masks from the dataset. (a) DRIVE. (b) CHASE-DB1. (c) STARE. }
	\label{fig_mask}
\end{figure}

\begin{figure*}[!t]
	\setlength{\fboxsep}{0pt}%
	\setlength{\fboxrule}{0.2pt}%
	
	\centering
	\subfloat[]{\includegraphics[width=0.3\textwidth]{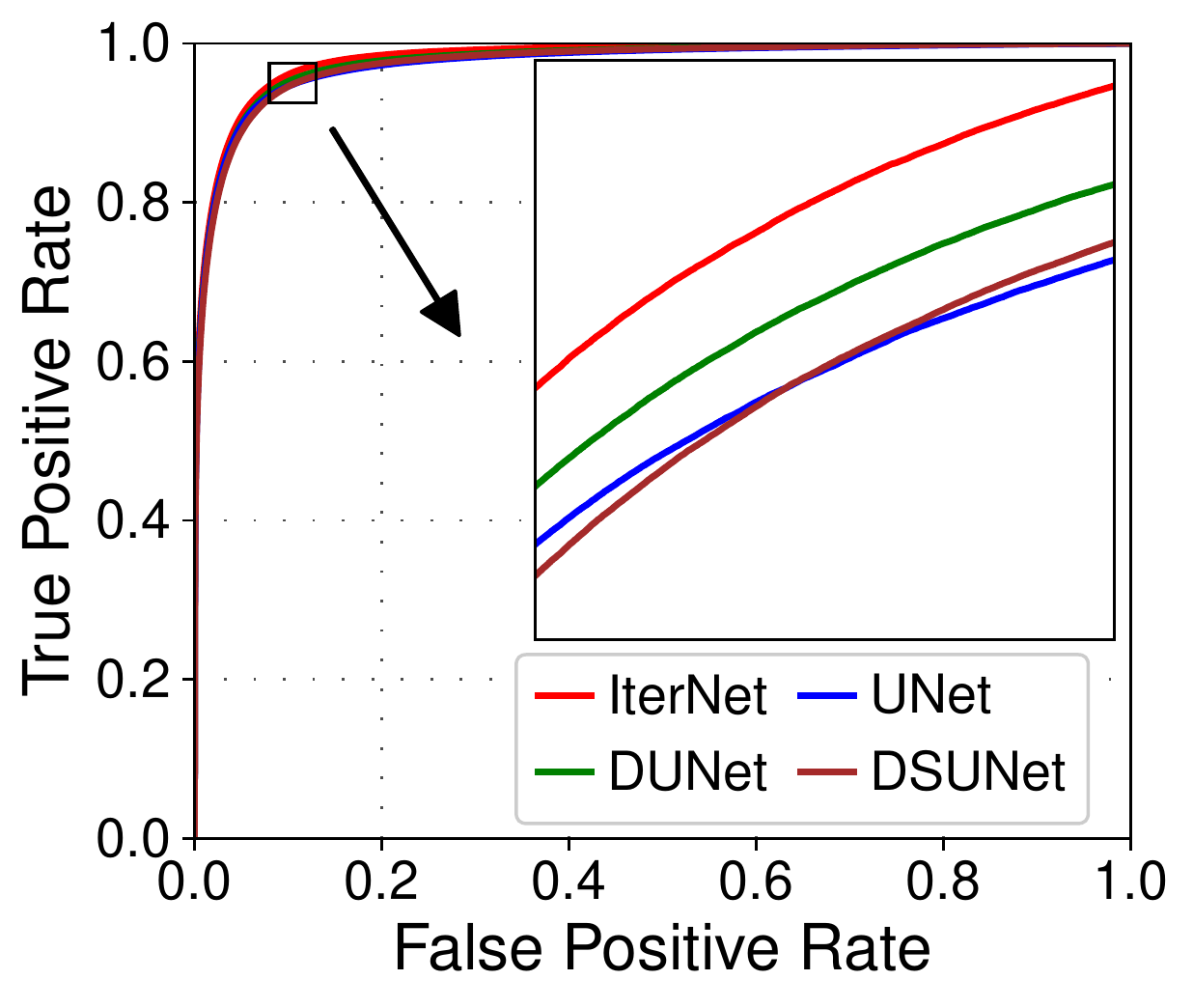}}
	\hfil
	\subfloat[]{\includegraphics[width=0.3\textwidth]{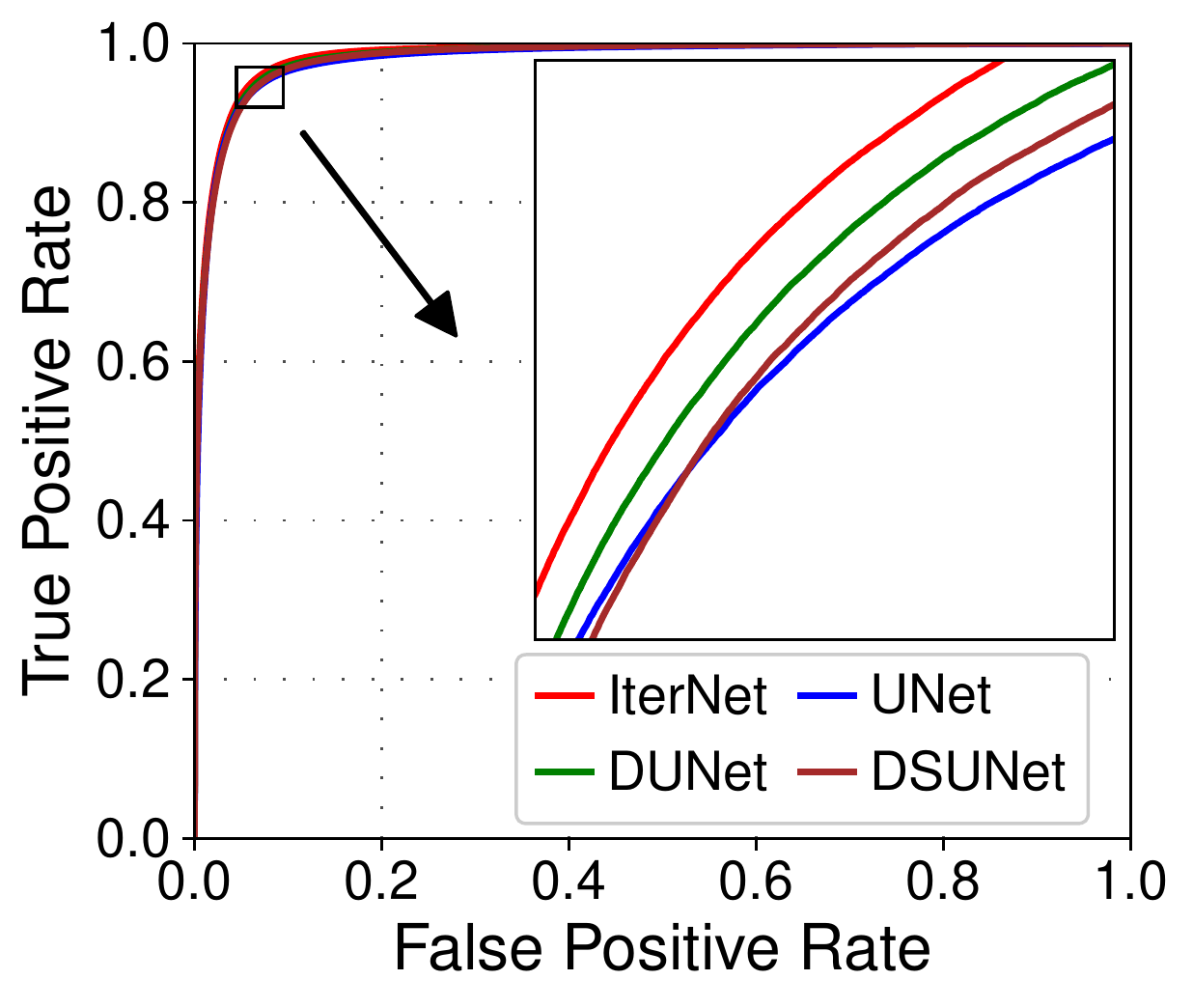}}
	\hfil
	\subfloat[]{\includegraphics[width=0.3\textwidth]{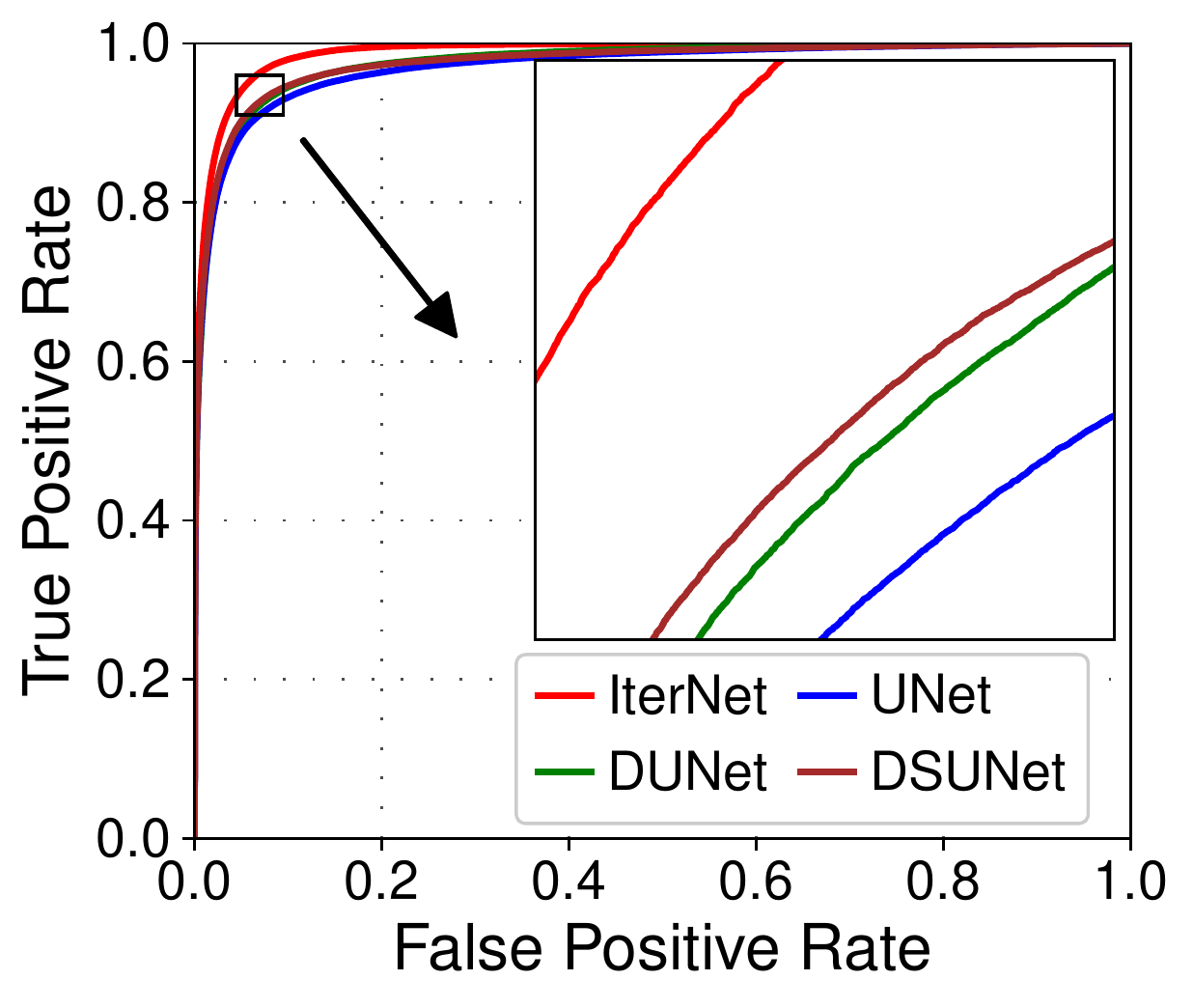}}
	\caption{ROC Curves on Three Datasets (With Masks). (a) DRIVE. (b) CHASE-DB1. (c) STARE.}
	\label{fig_auc}
\end{figure*}

\section{Experiments}\label{section_experiment}

In this section, we will give a detailed description of the experimental design, the results, the comparisons, and discussion on advantages and shortcomings of IterNet. All the experiments are performed on a GPU server, which has four NVIDIA Tesla V100 SXM2 GPU with 32GB memory each, and two Intel Xeon Gold 5122 CPU. For each model, we only use one GPU for fair comparison. The number of the iteration of the mini-UNet is set to three ($N=4$ in Fig.~\ref{fig_structure}), as we find that, for these three datasets, larger numbers only bring a minor improvement on the performance at the cost of much longer training and prediction times.

We used three popular datasets, \ie, DRIVE \cite{staal:2004-855}, CHASE-DB1 \cite{owen2009measuring}, and STARE \cite{845178}, as shown in the first row of Fig.~\ref{fig_mask}, in our experiments. They are all in different formats and different image sizes. The images are respectively in \texttt{.tif} (565$\times$584), \texttt{.jpg} (999$\times$960), and \texttt{.ppm} (700$\times$605). We train three different IterNet models with these three datasets. Because UNet does not require to fix the size of input images, we can use the same model configuration for all the three models. Also the model can take as input whole images or image patches. For training, we randomly extracted image patches from the images. For prediction, overlapping image patches are extracted with the stride of 3 (we compared the stride of 3 and of 8 in the supplementary material), and we used the average of all overlapping image patches as the prediction.

\begin{figure*}[!t]
	\setlength{\fboxsep}{0pt}%
	\setlength{\fboxrule}{0.2pt}%
	
	\centering
	\includegraphics[width=0.95\textwidth, height=0.875\textwidth]{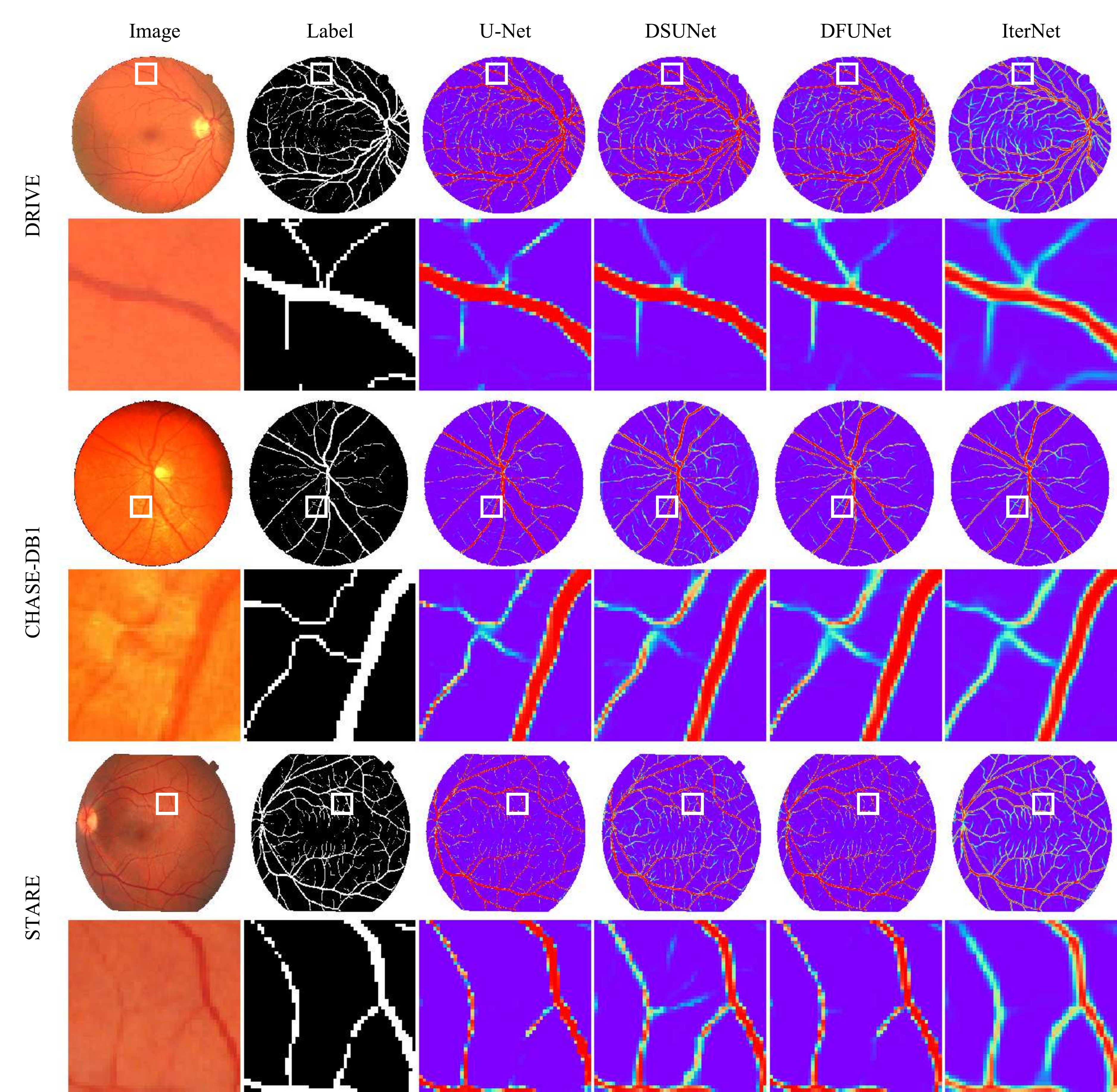}
	\caption{Visualization of the segmentation results on DRIVE, CHASE-DB1, and STARE datasets.}
	\label{fig_vis}
\end{figure*}

The second row of Fig.~\ref{fig_mask} is the field of view (FoV) masks of the retinal images. Although the DRIVE dataset provides official masks for the test images, the other two datasets do not have such masks. In order to assure fair comparison, we also generated the FoV masks for CHASE-DB1 and STARE. This can be easily done by simple color thresholding on the raw images because the pixels out of FoV are usually close to black.
In DRIVE dataset, we use 20 images for training and 20 images for testing. In CHASE-DB1 and STARE, there is no official description about training and test splits, so we made 20 and 16 images, respectively, as the training sets, and the remaining 8 and 4 images as the test sets. We do not use any validation images.

We compared our models with some state-of-the-art ones, including UNet \cite{UNet}, DenseBlock-UNet \cite{8697107, 8379359}, and Deform-UNet \cite{JIN2019}. We trained and evaluated these models using their public code by ourselves on three datasets, because training and test splits are unknown for CHASE-DB1 and STARE and we want to produce the receiver operating characteristics (ROC) curves, which are presented in Fig.~\ref{fig_auc}. As can be seen in the figure, in most cases, IterNet shows better performance than the other three models. The performance boost is small as all state-of-the-art models already have good performance (AUC $>$ 0.97). Yet, the results proved that our model gave a stable performance over all three datasets. Among them, IterNet worked well on the STARE dataset, which has fewer training and test samples. This result implies that the IterNet can find the proper features and patterns in the vessel network even with limited training images. In contrast, all other models suffer from a big deterioration in the STARE dataset. Among the other three state-of-the-art models, Deform-UNet usually showed significantly better performance due to its dynamic receptive field. However the STARE dataset decreased its advantages over the DenseBlock-UNet because the dense-block module makes the model less prone to overfitting.

We also compared the results with some existing models, including the aforementioned three UNet-based models, %
Residual UNet \cite{alom2018recurrent}, Recurrent UNet\cite{alom2018recurrent}, R2UNet \cite{alom2018recurrent}, and one iterate prediction methods, \ie Iter-Seg \cite{7042289}%
, which have been introduced in Section~\ref{Related_Works_section}. Only the results of %
UNet, DenseBlock-UNet, and Deform-UNet were from our reproduced tests, while all other results were adopted from the corresponding papers. The results on the DRIVE dataset are shown in Table~\ref{table_detect_performance_drive}. We show results of two variants of IterNet. Both of them use image patches with the size of 128 for training. In prediction, one takes a whole image as input and outputs the final results, while the other (denoted by ``patched'') uses image patches for both training and prediction, and the resulting vessel maps are concatenated. As we can see, image patch-based prediction brought some improvement while it costs longer running time (see the supplementary material for the detailed time cost). These two variants have shown the superior AUCs to all other models. Actually, they are the only models in our test that have AUCs higher than $0.98$.

We also conducted the comparison experiments on CHASE-DB1 and STARE, which do not have officially-specified training and test sets as well as the FoV masks. Therefore, we only compare the results with the results of our reproduced models with the same settings. The results are shown in Tables \ref{table_detect_performance_chase} and \ref{table_detect_performance_stare}. To ensure a fair comparison, we list the performance both with or without FoV masks.
It can be seen that the proposed IterNet has the best performance in the most metrics on both datasets.

However, all the metrics above are pixel-level and do not reflect the segmentation performance on the vessel network level. Therefore, we adopt a new metric, \ie, \textit{connectivity} \cite{MOCCIA201871,6019055}, which is an important requirement for clinicians to conduct analysis on retinal images using some vessel-related patterns, such as crossing or branching \cite{kawasaki}. The connectivity $C$ is defined as follows.
\begin{equation}
    C(\theta) =\left\{
	\begin{aligned}
	&1 - \frac{|S_\text{P}(\theta)-S_\text{G}|}{S_\text{Max}} & & \text{for }  |S_{\text{P}}(\theta)-S_\text{G}|\leq S_\text{Max}\\
	&0 & & \text{otherwise} %
	\end{aligned}
	\right.
\end{equation}
where $S_\text{P}(\theta)$ is the number of segments in the predicted segmentation binarized with threshold $\theta$, and $S_\text{G}$ the number of segments in the gold standard segmentation, respectively. $S_\text{Max}$ is the maximum number of segments allowed for one vessel map. Since the maximum number of segments involves the total vessel length $L$, it should be defined according to $L$, which can be calculated by skeletonizing the gold standard and counting the number of skeleton pixels. We set $S_\text{Max}=\alpha L$ and we make $\alpha=0.05$ in this experiment. With this definition, we drew a curve of $\theta$ versus $C(\theta)$ (refer to the supplementary material for some examples). We adopt the area under this curve as connectivity metric (abbreviated to \emph{Conn}.).
As shown in Tables \ref{table_detect_performance_drive}, \ref{table_detect_performance_chase}, and \ref{table_detect_performance_stare}, IterNet achieved the highest connectivity in all three datasets.

\begin{table*}[!t]
	\caption{Performance comparison on the DRIVE dataset (with mask).}
	\label{table_detect_performance_drive}
	\centering
	\begin{tabular}{l|c|cccccc}
		\hline
		Method & Year & Conn. & F1 Score & Sensitivity & Specificity & Accuracy & AUC\\ 
		\hline
		Iter-Seg \cite{7042289} &2016&-&- & 0.739 & 0.978 & 0.949& 0.967\\
		UNet(reported \cite{JIN2019}) & 2018& 0.7948 &  0.8174(0.8021)& 0.7822(-)& 0.9808(-) &0.9555(0.9681)  &0.9752(0.9830)\\		
		Residual UNet \cite{alom2018recurrent}&2018&-& 0.8149& 0.7726& 0.9820& 0.9553& 0.9779\\
		Recurrent UNet\cite{alom2018recurrent}& 2018&-& 0.8155& 0.7751& 0.9816 &0.9556& 0.9782\\
		R2UNet \cite{alom2018recurrent}&2018&-& 0.8171 &0.7792& 0.9813& 0.9556 &0.9784\\
		DenseBlock-UNet& 2018& 0.8332 & 0.8146 &0.7928& 0.9776&  0.9541&0.9756\\
		DUNet(reported \cite{JIN2019}) &2019& 0.8314 &0.8190(0.8203)&  \textbf{0.7863}(-) & 0.9805(-) &  0.9558(0.9697)& 0.9778(0.9856)\\
		
		\textbf{IterNet} &2019& 0.9001& \textbf{0.8218} & 0.7791 &0.9831 & \textbf{0.9574}& 0.9813\\
		\textbf{IterNet(Patched)} & 2019& \textbf{0.9193} &0.8205 & 0.7735 & 0.9838 &0.9573& \textbf{0.9816}\\
		
		\hline
	\end{tabular}
\end{table*}

\begin{table*}[!t]
	\caption{Performance comparison on the CHASE-DB1 dataset.}
	\label{table_detect_performance_chase}
	\centering
	\begin{tabular}{cl|c|cccccc}
		\hline
		FoV &Method & Year & Conn. & F1 Score & Sensitivity & Specificity & Accuracy & AUC\\ 
		\hline		
	
		\multirow{4}{*}{Without Masks} & UNet & 2018& 0.8198 &0.7993&   0.7840& 0.9880 & 0.9752 & 0.9870\\		
		&DenseBlock-UNet& 2018 & 0.8269 & 0.8005 &0.8177& 0.9848&  0.9743&0.9880\\
		&DUNet &2019& 0.8402 &0.8000&  0.7858 & 0.9880 &  0.9752& 0.9887\\
		&\textbf{IterNet} & 2019& \textbf{0.9091} &  \textbf{0.8072} &  \textbf{0.7969} & \textbf{0.9881} & \textbf{0.9760}& \textbf{0.9899}\\
		
		\hline
		
		\multirow{4}{*}{With Masks}&UNet & 2018& 0.8198 &0.7993&   0.7841& \textbf{0.9823} & 0.9643 & 0.9812\\
		&DenseBlock-UNet& 2018& 0.8269 & 0.8006 &0.8178&0.9775 &  0.9631&0.9826\\
		&DUNet &2019& 0.8402 &0.8001 &  0.7859 & 0.9822 &  0.9644& 0.9834\\
		&\textbf{IterNet} & 2019& \textbf{0.9091} &  \textbf{0.8073} &  \textbf{0.7970} & \textbf{0.9823} & \textbf{0.9655}& \textbf{0.9851}\\
		
		\hline
	\end{tabular}
\end{table*}

\begin{table*}[!t]
	\caption{Performance comparison on the STARE dataset.}
	\label{table_detect_performance_stare}
	\centering
	\begin{tabular}{cl|c|cccccc}
		\hline
		FoV &Method & Year & Conn. & F1 Score & Sensitivity & Specificity & Accuracy & AUC\\ 
		\hline					
		
		\multirow{4}{*}{Without Masks} &UNet  & 2018& 0.7148 &0.7594&   0.6681& 0.9939 & 0.9736 & 0.9779\\
		&DenseBlock-UNet& 2018& 0.7229 & 0.7691 &0.6807& 0.9940& 0.9745 &0.9801\\
		&DUNet &2019& 0.7479 &0.7629&  0.6810 & \textbf{0.9931} &  0.9736& 0.9823\\
		&\textbf{IterNet} & 2019&\textbf{0.8977}&  \textbf{0.8146} &  \textbf{0.7715} & 0.9919 & \textbf{0.9782}& \textbf{0.9915}\\
		
		\hline
		
		\multirow{4}{*}{With Masks}&UNet & 2018& 0.7148 &0.7595&   0.6681& 0.9915 & 0.9639 & 0.9710\\		
		&DenseBlock-UNet& 2018& 0.7229 & 0.7691 &0.6807& 0.9916& 0.9651 &0.9755\\
		&DUNet &2019& 0.7479 &0.7629&  0.6810 & \textbf{0.9903} &  0.9639& 0.9758\\		
		&\textbf{IterNet} & 2019&\textbf{0.8977}&  \textbf{0.8146} &  \textbf{0.7715} & 0.9886 & \textbf{0.9701}& \textbf{0.9881}\\
		\hline
	\end{tabular}
\end{table*}

We present some example results in Fig.~\ref{fig_vis}. As we can see, over all three datasets, our IterNet model worked the best. We consider that this is due to deep understanding of vessel networks by IterNet's iterative architecture: It knows how to connect vessel segments together even they look visually disconnected on the raw retinal images.

As introduced in Section~\ref{section_method}, weight-sharing among mini-UNets helps to avoid overfitting in the training process. We conduct an experimental test to see the actual performance of IterNet without weight-sharing. When $N=1$, there is no mini-UNets, the IterNet can be trained as common UNet; when $N=2$, the mini-UNet only runs for one time and we get an AUC of 0.9795 on the DRIVE dataset, which is very similar with the performance of \texttt{Out1} from the IterNet with $N=3$; while when $N\ge2$, IterNet encounters serious overfitting problems that the loss can reach a low level on the training set while keeps high on the test set. 
We also conduct an experiment to test the performance of IterNet without skip connection, the AUCs respectively drop to 0.9799, 0.9770, 0.9808 on three datasets (refer to the supplementary material for more results).

\section{Conclusion}\label{conclusion_section}

In this paper, we propose a segmentation model named IterNet to address some existing problems in retinal image segmentation. We use a standard UNet to analyze the raw input images and map them into an initial prediction of the vessel network. In order to remove errors, such as inconsistent vessels, missing pixels, \etc, which are very common in existing vessel segmentation models, we add an iteration of mini-UNets after UNet, and use the output of UNet as the input of the following mini-UNets. By introducing weight-sharing in mini-UNets and skip-connections, we successfully empower IterNet with the ability to find possible defections in the intermediate results and fix them in a reasonable way. The experimental results prove that the proposed IterNet has achieved state-of-the-art performance over three commonly-used datasets.

\section*{Acknowledgment}
This work was supported by Council for Science, Technology and Innovation (CSTI), cross-ministerial Strategic Innovation Promotion Program (SIP), ``Innovative AI Hospital System'' (Funding Agency: National Institute of Biomedical Innovation, Health and Nutrition (NIBIOHN)).

{\small

}

\pagebreak
\clearpage

\begin{strip}
\begin{center}
	 {\Large \bf IterNet: Retinal Image Segmentation\\Utilizing Structural Redundancy in Vessel Networks\\(Supplemental Materials) \par}
	
	 \vskip .5em
	 \vspace*{12pt}
\end{center}
\vspace{0.5 in}
\end{strip}
\setcounter{equation}{0}
\setcounter{figure}{0}
\setcounter{table}{0}
\setcounter{page}{1}
\makeatletter

\begin{figure*}
	\setlength{\fboxsep}{0pt}%
	\setlength{\fboxrule}{0.2pt}%
	
	\centering
	\subfloat[]{\includegraphics[width=0.3\textwidth]{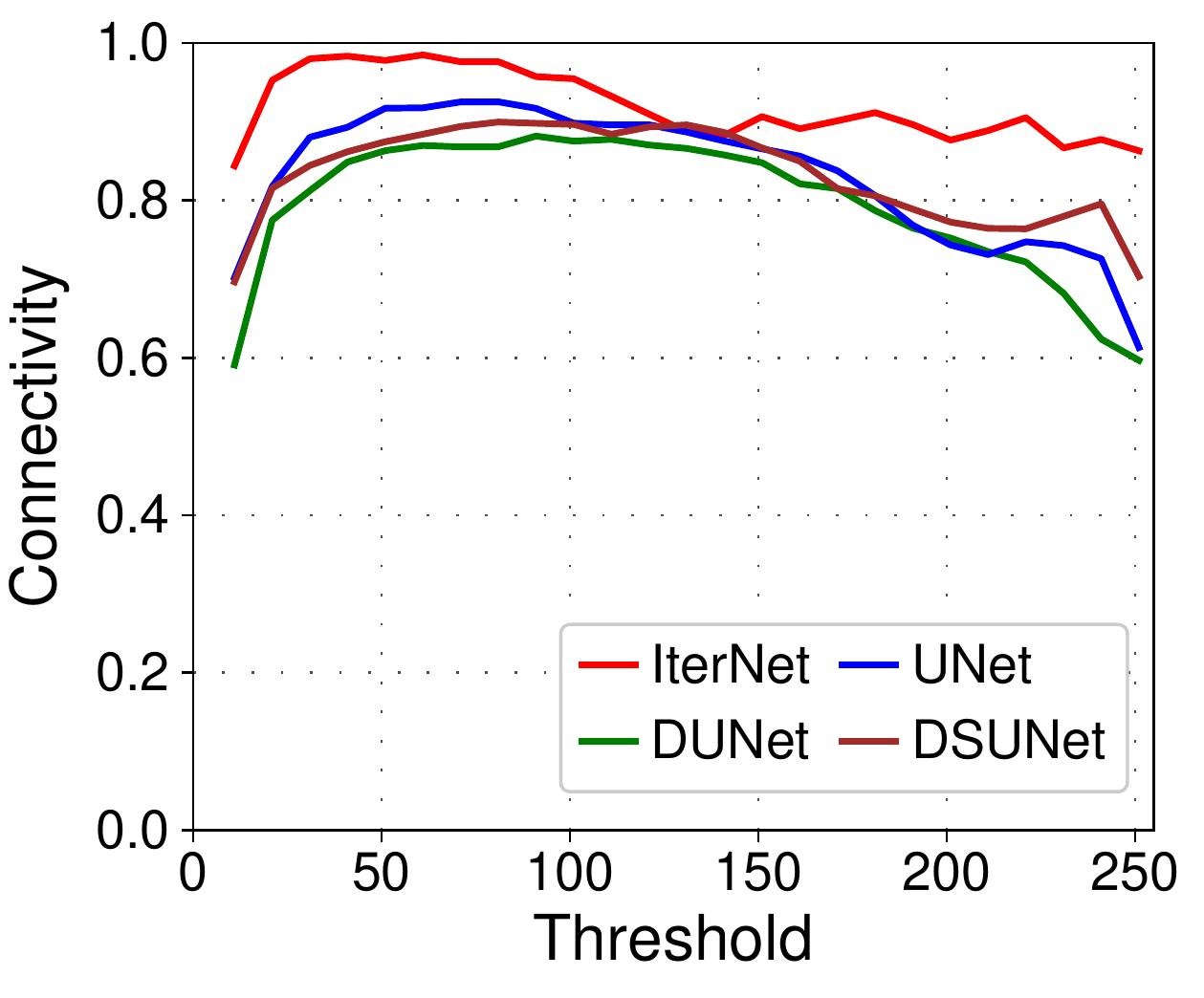}}
	\hfil
	\subfloat[]{\includegraphics[width=0.3\textwidth]{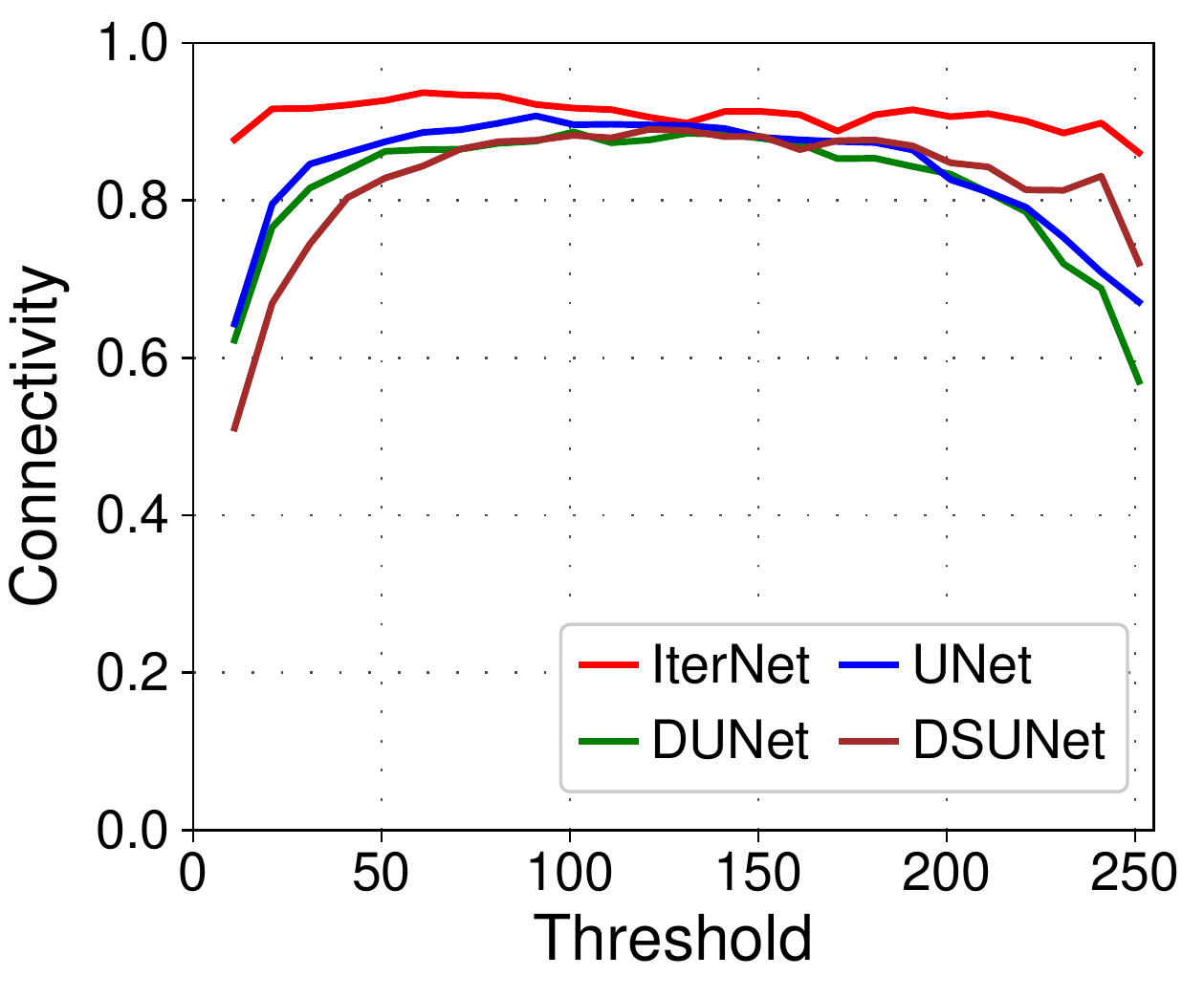}}
	\hfil
	\subfloat[]{\includegraphics[width=0.3\textwidth]{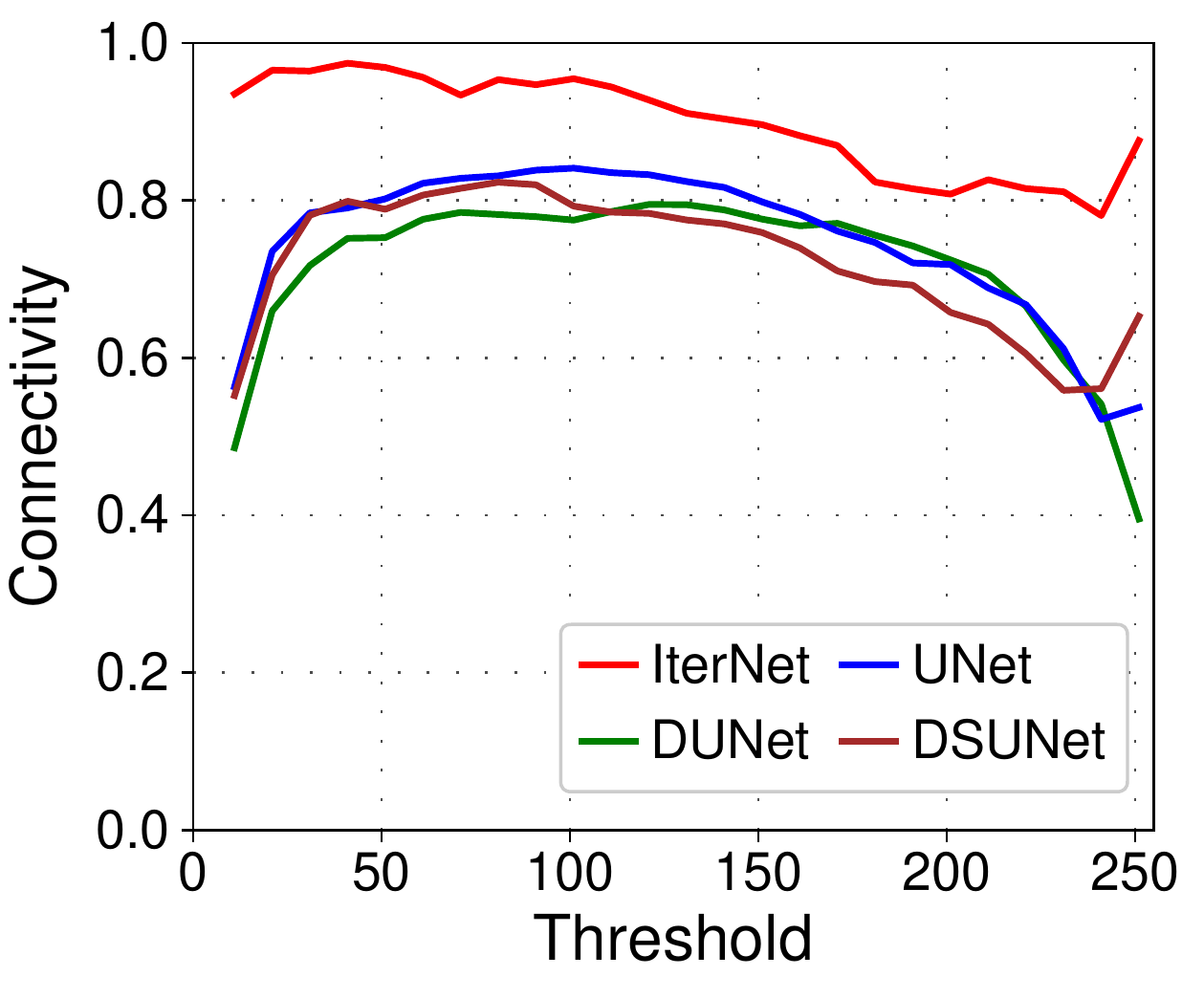}}
	\caption{Connectivity versus threshold on the three datasets: (a) DRIVE. (b) CHASE-DB1. (c) STARE.}
	\label{fig_conn}
\end{figure*}

Figure \ref{fig_conn} shows the connectivity value under different threshold for several methods on three popular datasets, \ie, DRIVE \cite{Sstaal:2004-855}, CHASE-DB1 \cite{Sowen2009measuring}, and STARE \cite{S845178}. The area under the curve is used as the measurement in the paper. We can see that IterNet almost always outperforms the other three methods.

Tables~\ref{table_detect_performance_drive1}, \ref{table_detect_performance_chase1}, and \ref{table_detect_performance_stare1} give the results on various criteria for two variants of IterNet. The first one is the IterNet model without skip connections among the first layer of the base UNet and the first layers of the mini-UNets, while the second one is to replace mini-UNets in IterNet with full-size UNets. Results show that they both suffer from a performance drop on all three datasets.

Table~\ref{table_time_costs} shows the detailed time cost in the inference process. We used $128\times 128$ image patches and tested different strides (the image patches are extracted every 3 or 8 pixels in both horizontal and vertical directions). We can see that a smaller stride may lead to a better refinement, while it also brings much bigger time cost.

\begin{figure*}[!t]
	\setlength{\fboxsep}{0pt}%
	\setlength{\fboxrule}{0.2pt}%
	
	\centering
	\subfloat[]{\includegraphics[width=0.35\textwidth]{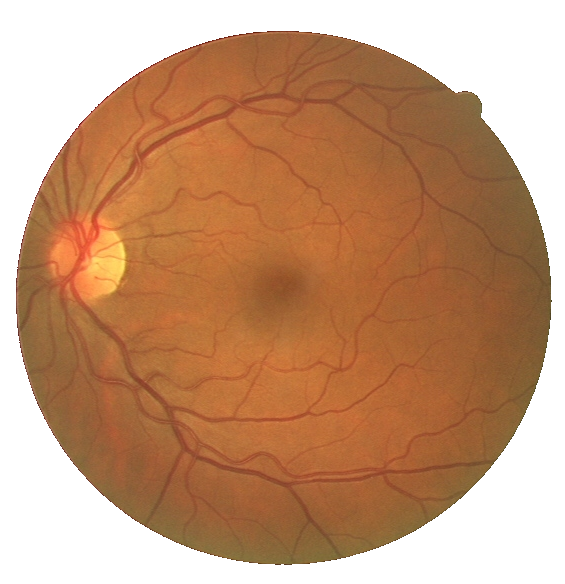}}
	\hfil
	\subfloat[]{\includegraphics[width=0.35\textwidth]{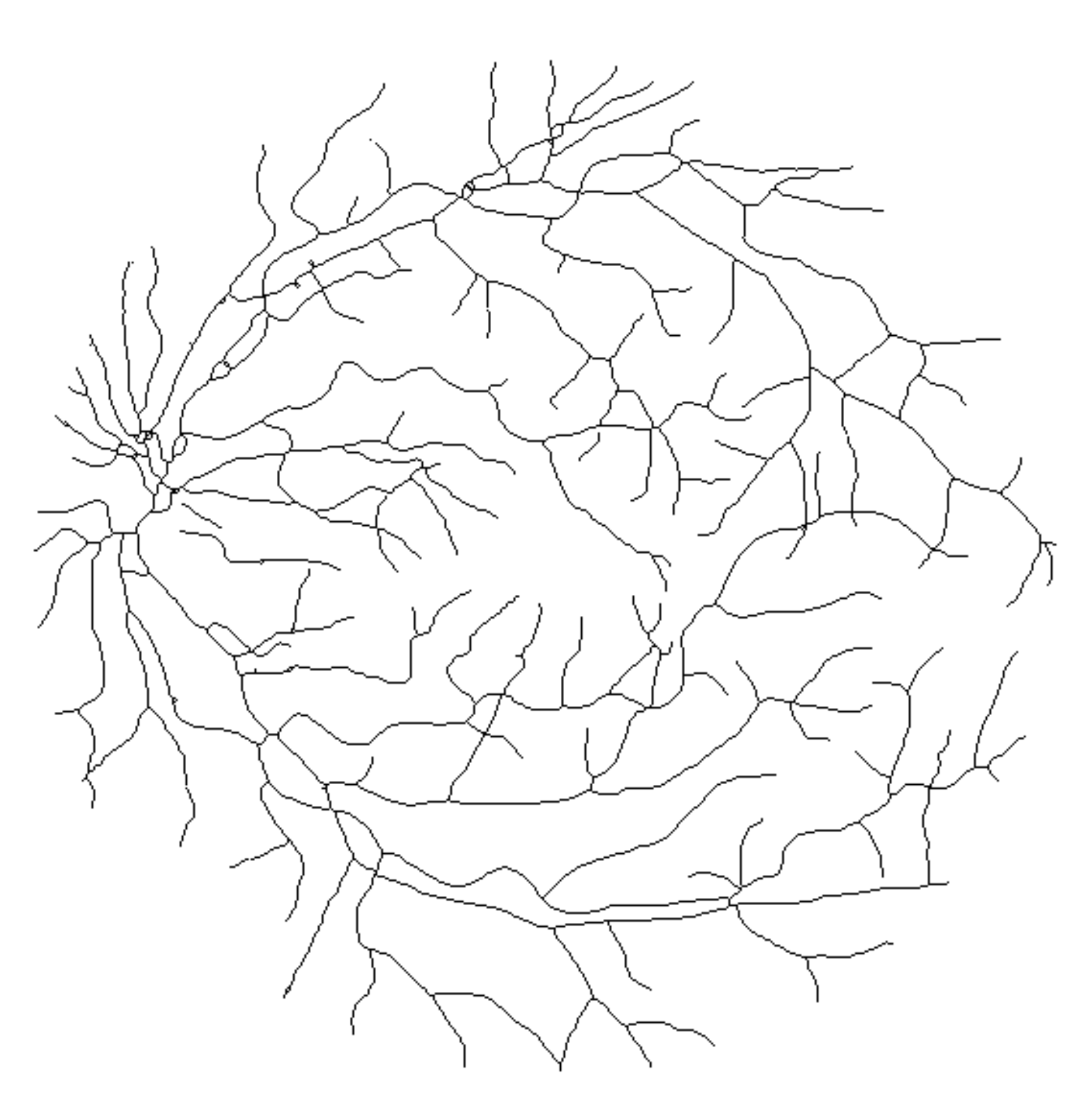}}

	\subfloat[]{\includegraphics[width=0.35\textwidth]{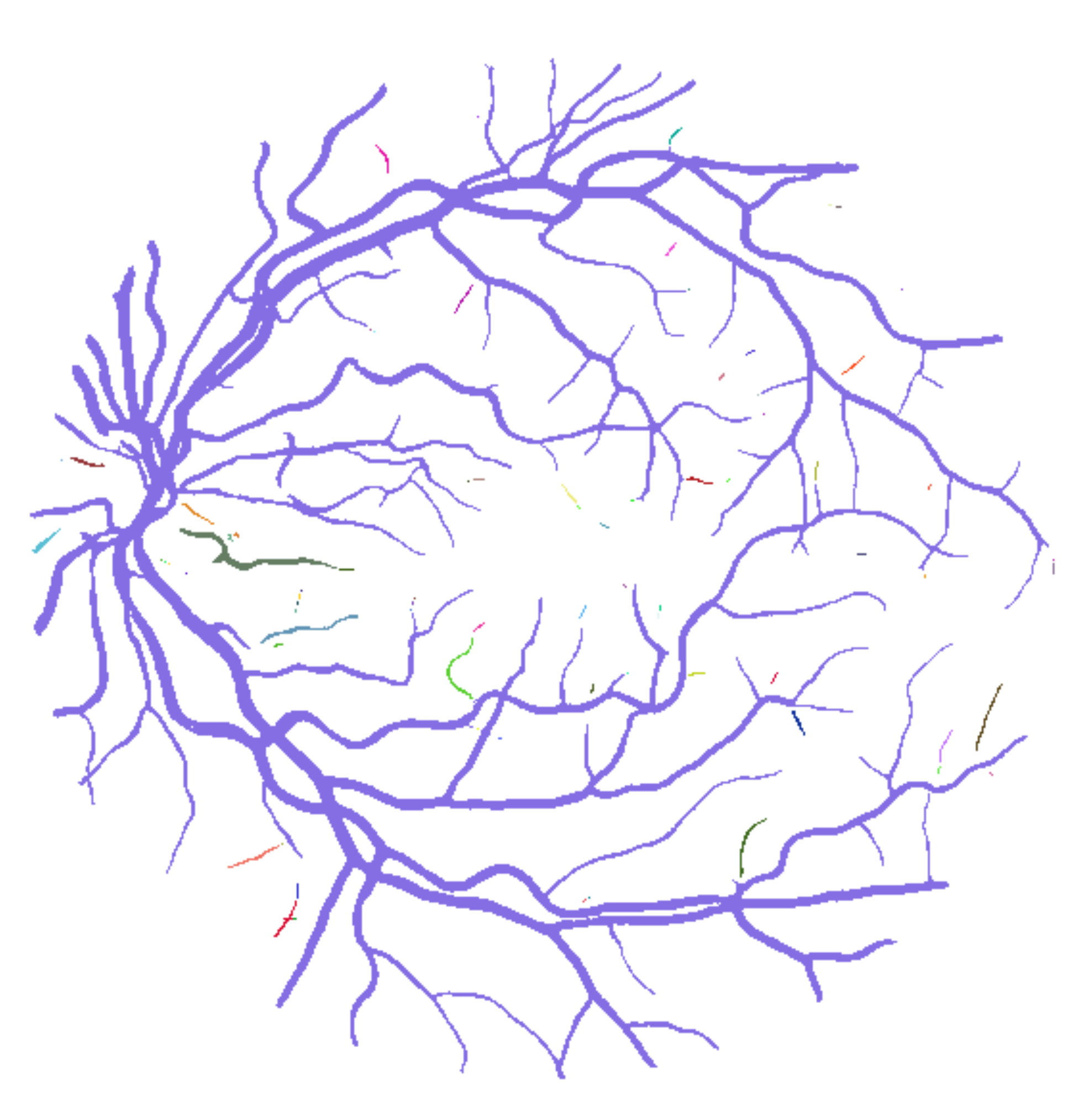}}
	\hfil
	\subfloat[]{\includegraphics[width=0.35\textwidth]{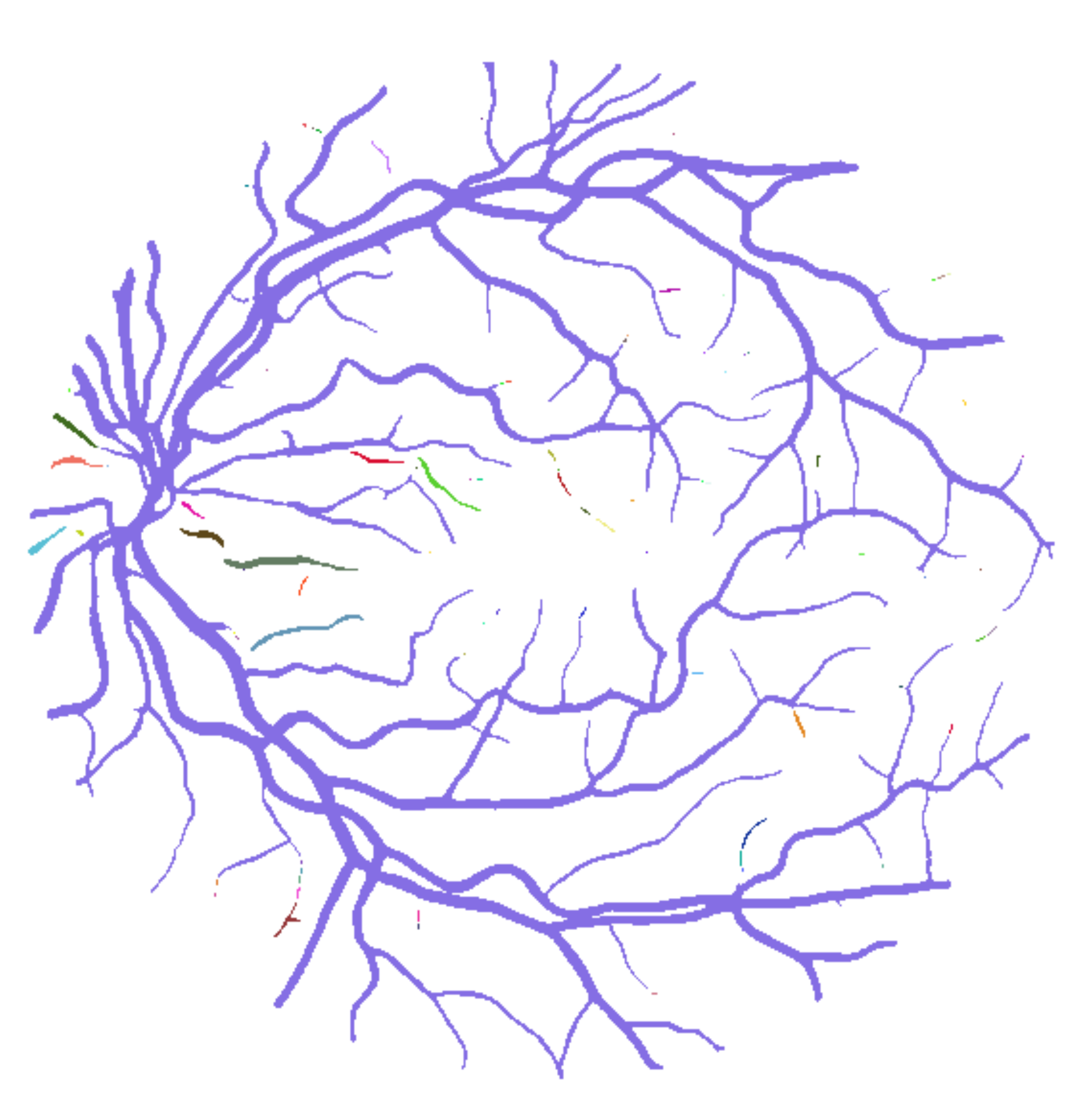}}

	\subfloat[]{\includegraphics[width=0.35\textwidth]{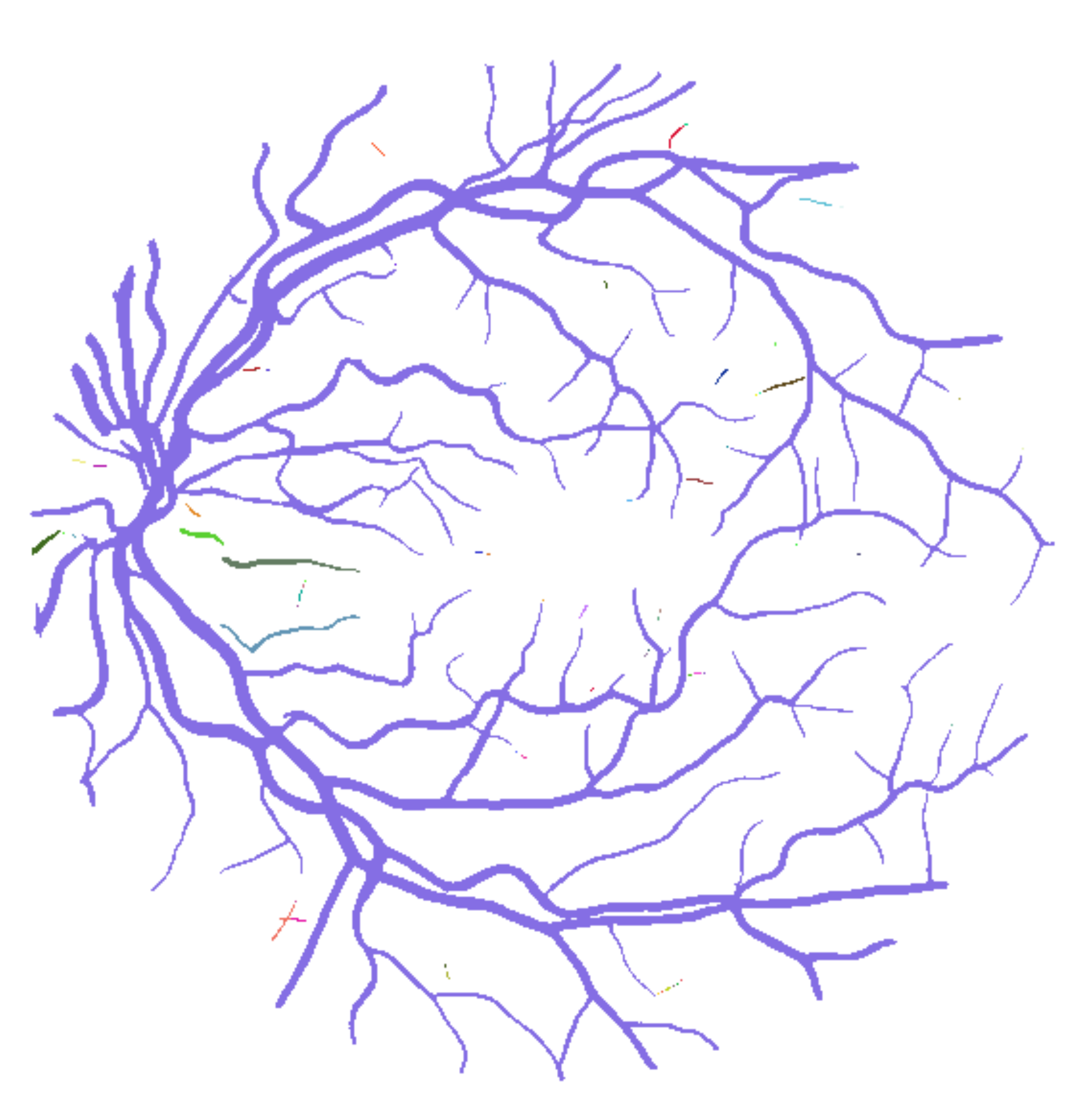}}
	\hfil
	\subfloat[]{\includegraphics[width=0.35\textwidth]{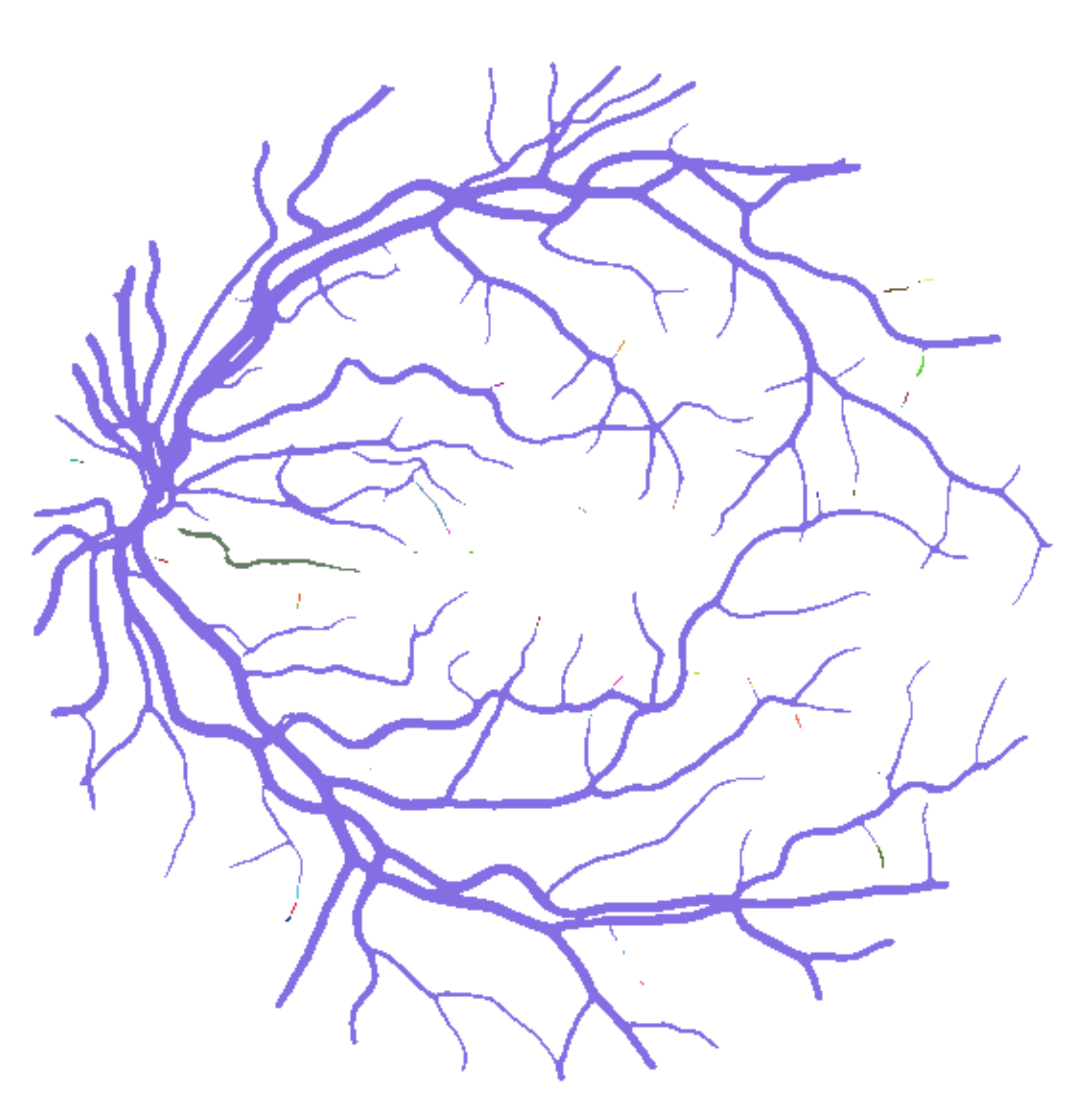}}

	\caption{Vessel segments visualization of a retina image from DRIVE (when $\text{threshold}=110$ and the connectivity values are provided for each method in the parentheses). (a) Raw image. (b) Extracted center-line from the ground-truth. (c) UNet (0.7905). (d) DenseNet (0.8282). (e) DUNet (0.8290). (f) IterNet (0.9049). Different colors means different segments. IterNet produces the fewest segments among all these methods.}
	\label{fig_drive}
\end{figure*}

\begin{figure*}[!t]
	\setlength{\fboxsep}{0pt}%
	\setlength{\fboxrule}{0.2pt}%
	
	\centering
	\subfloat[]{\includegraphics[width=0.35\textwidth]{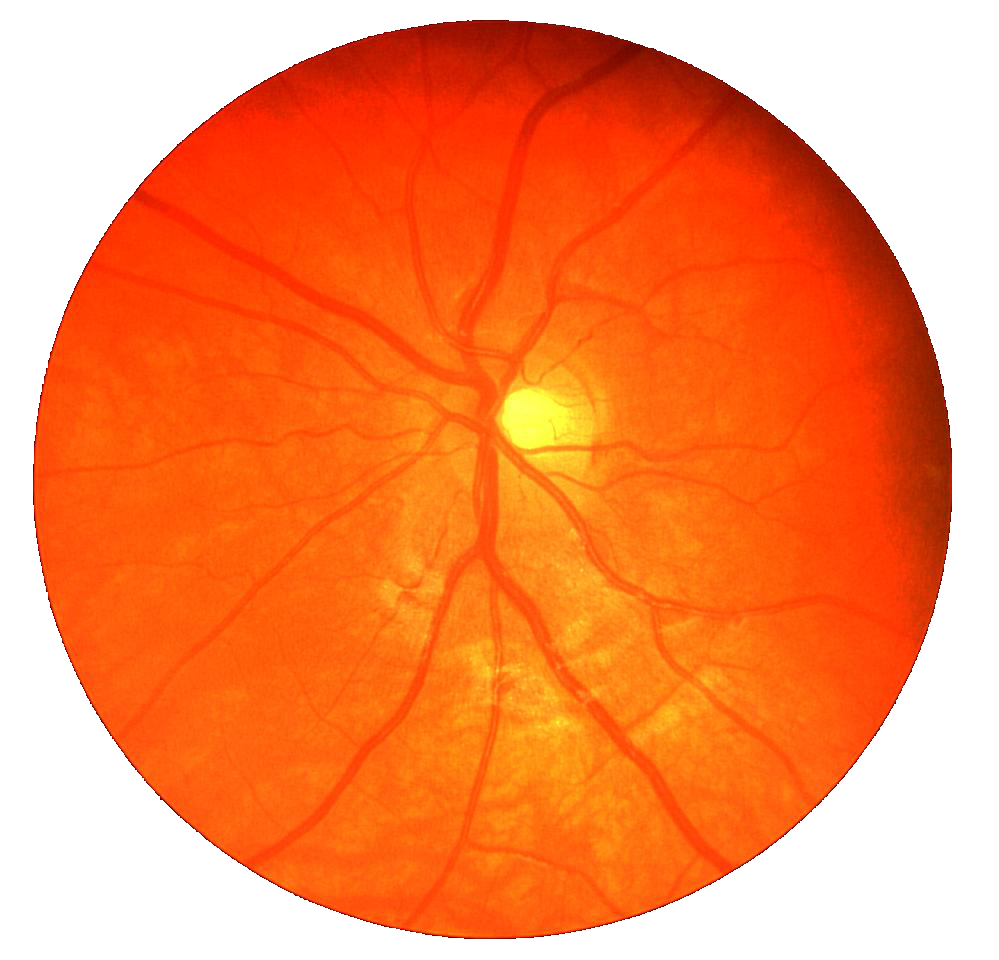}}
	\hfil
	\subfloat[]{\includegraphics[width=0.35\textwidth]{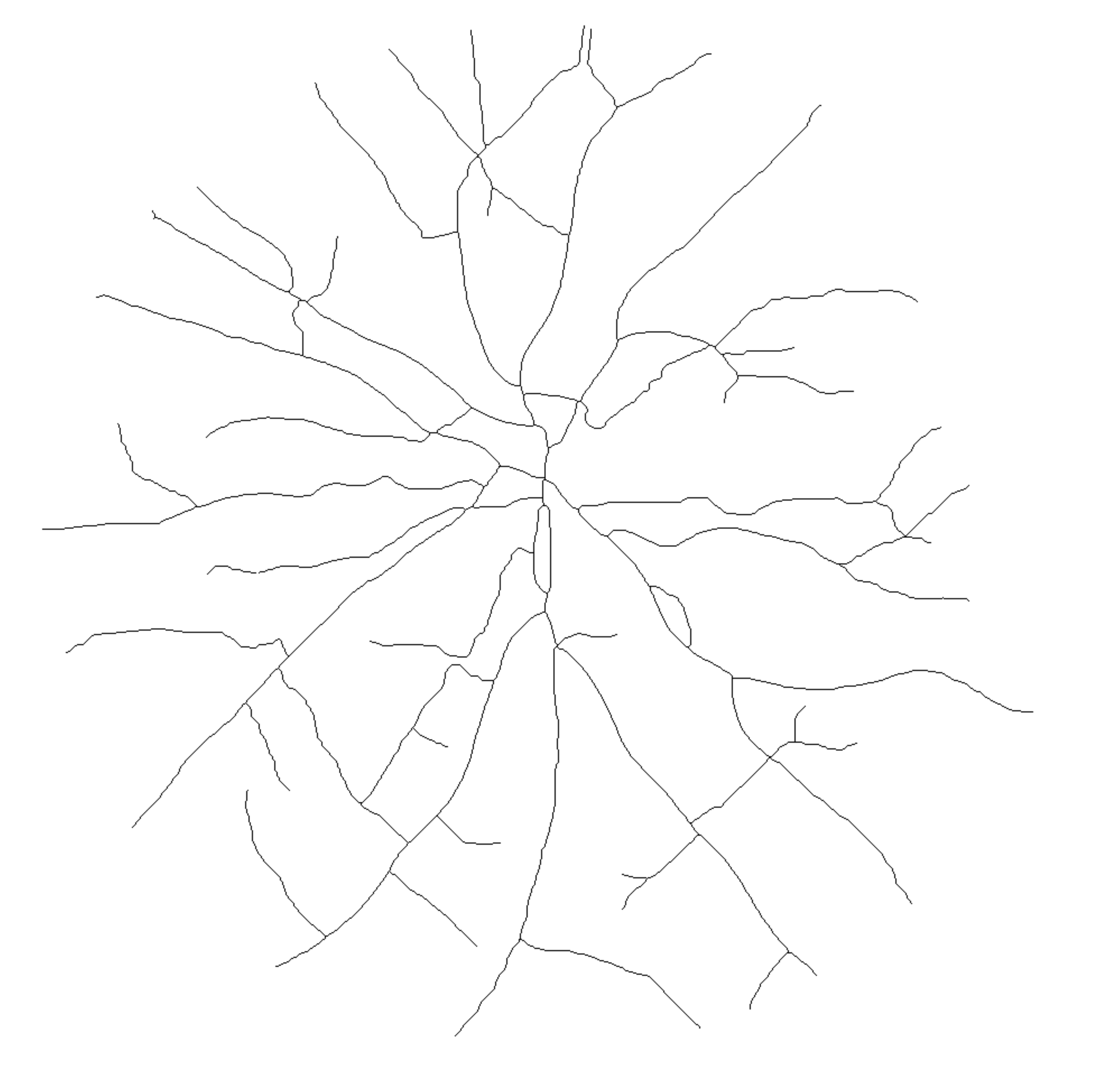}}

	\subfloat[]{\includegraphics[width=0.35\textwidth]{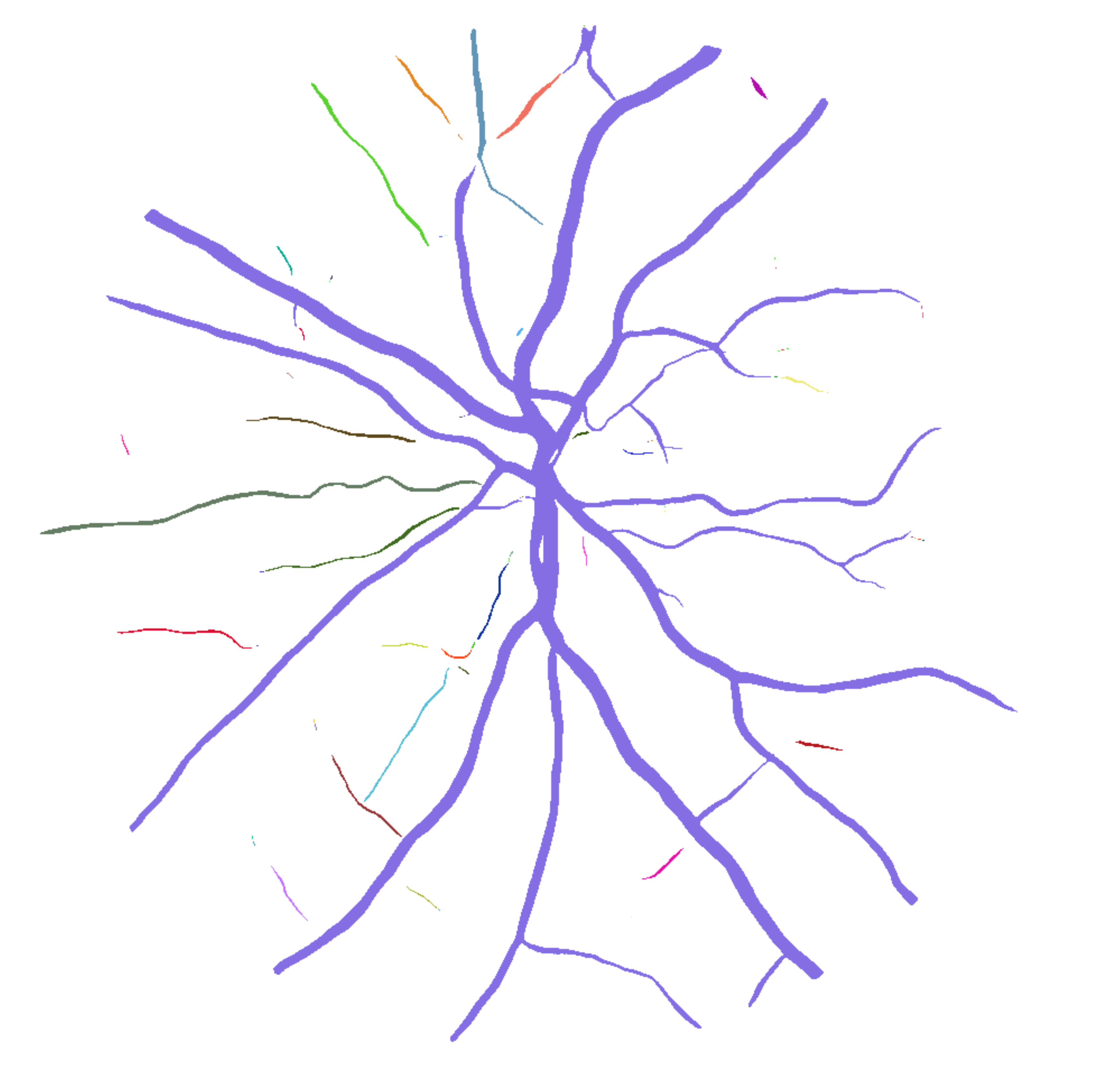}}
	\hfil
	\subfloat[]{\includegraphics[width=0.35\textwidth]{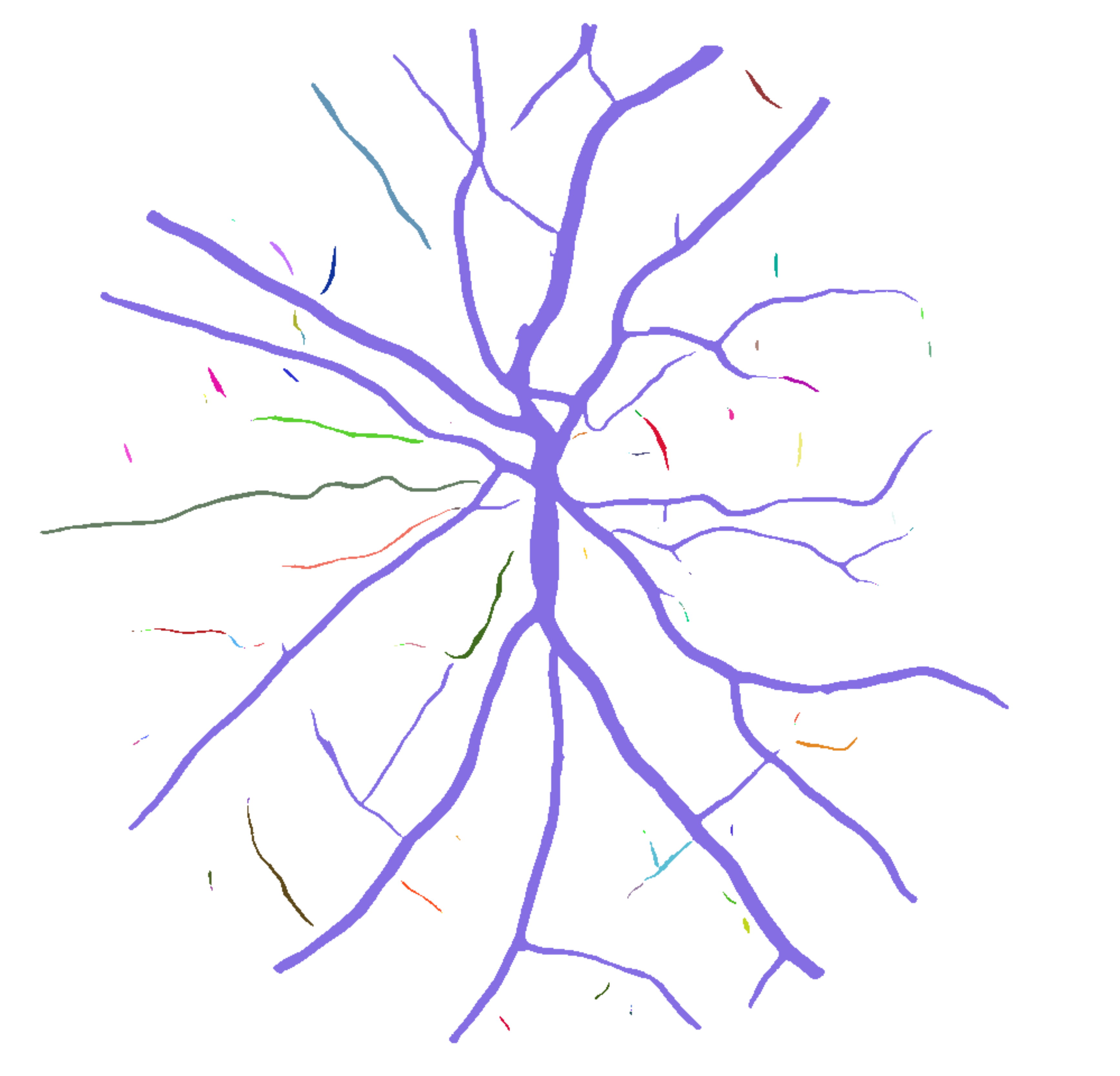}}

	\subfloat[]{\includegraphics[width=0.35\textwidth]{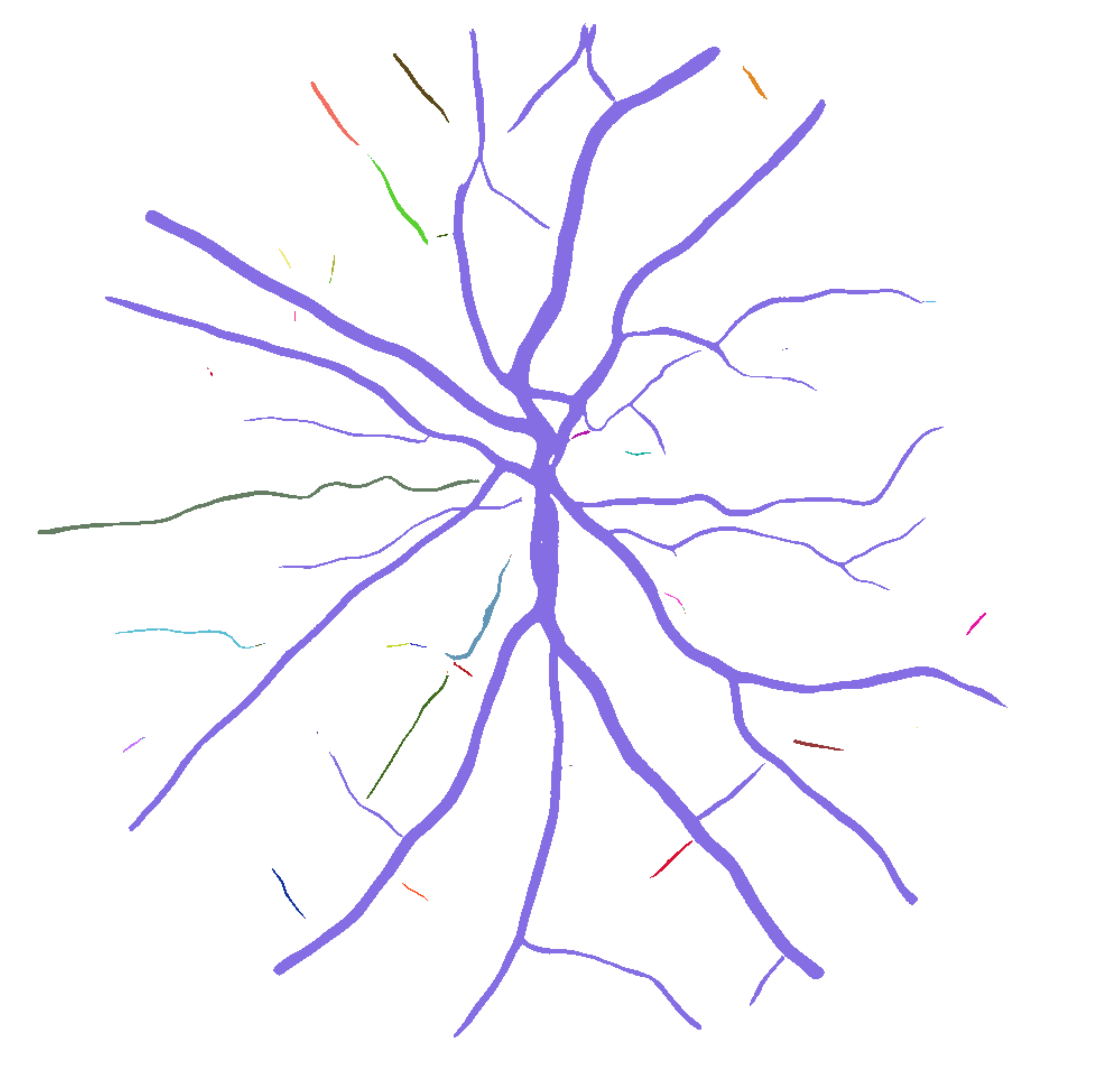}}
	\hfil
	\subfloat[]{\includegraphics[width=0.35\textwidth]{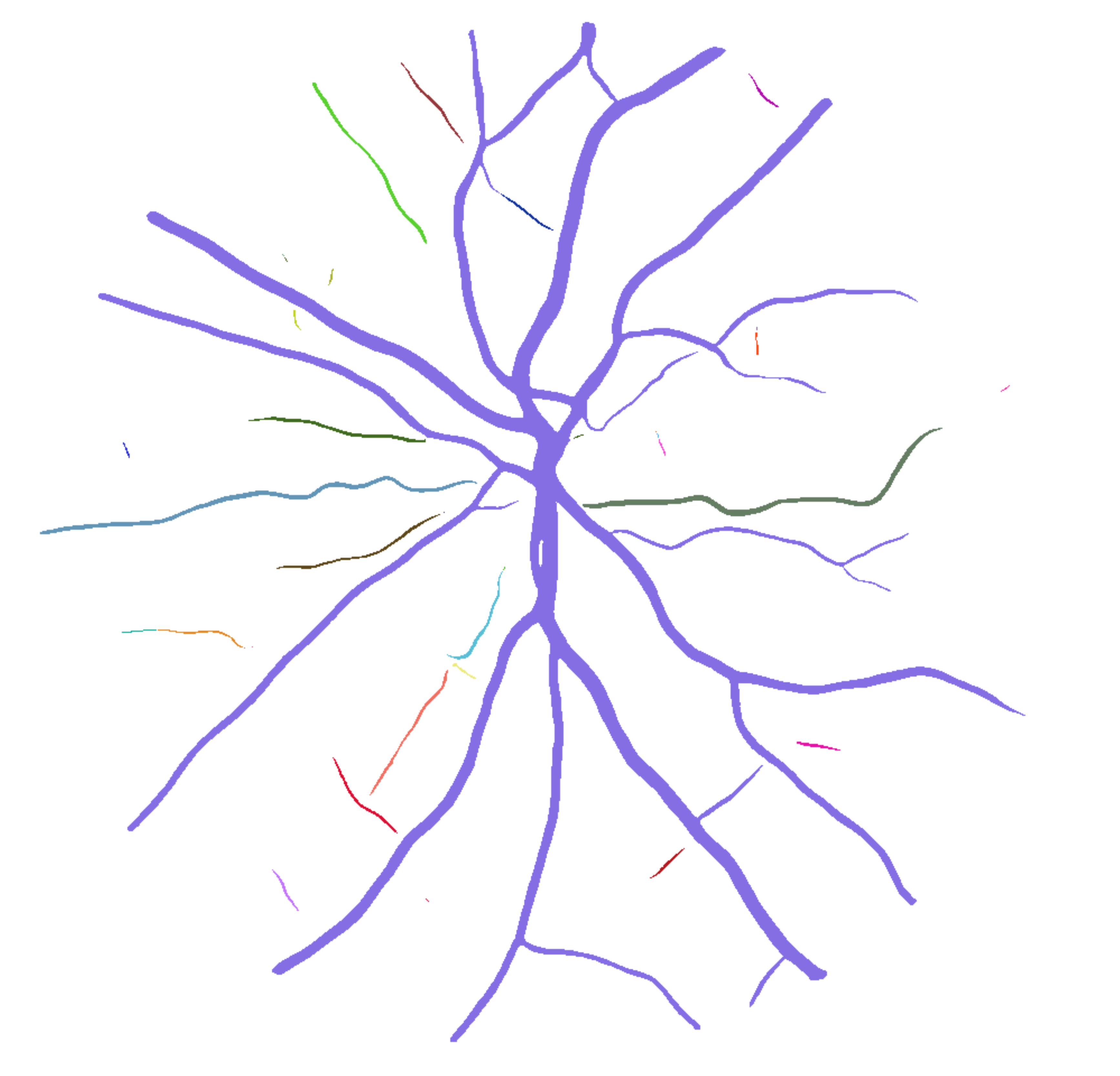}}

	\caption{Vessel segments visualization of a retina image from CHASE-DB1 (when $\text{threshold}=110$ and the connectivity values are provided for each method in the parentheses). (a) Raw image. (b) Extracted center-line from the ground-truth. (c) UNet (0.8085). (d) DenseNet (0.8019). (e) DUNet (0.8423). (f) IterNet (0.9034). IterNet also gives the smallest number of segments.}
	\label{fig_chasedb1}
\end{figure*}

\begin{figure*}[!t]
	\setlength{\fboxsep}{0pt}%
	\setlength{\fboxrule}{0.2pt}%
	
	\centering
	\subfloat[]{\includegraphics[width=0.35\textwidth]{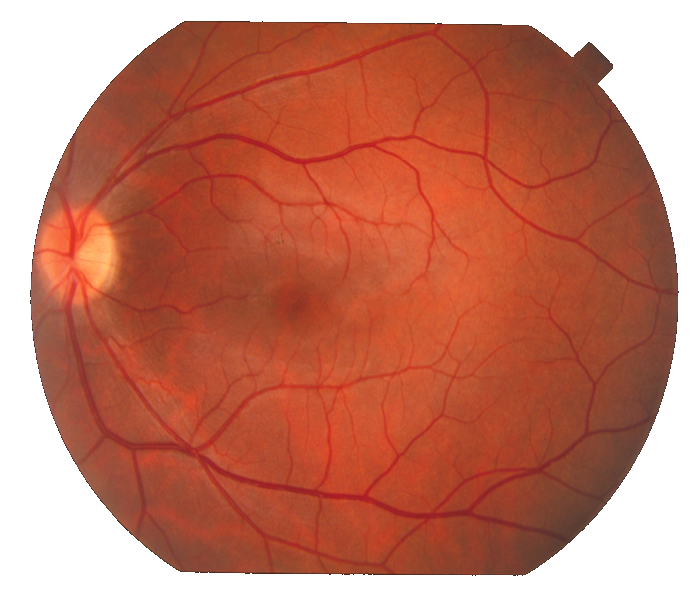}}
	\hfil
	\subfloat[]{\includegraphics[width=0.35\textwidth]{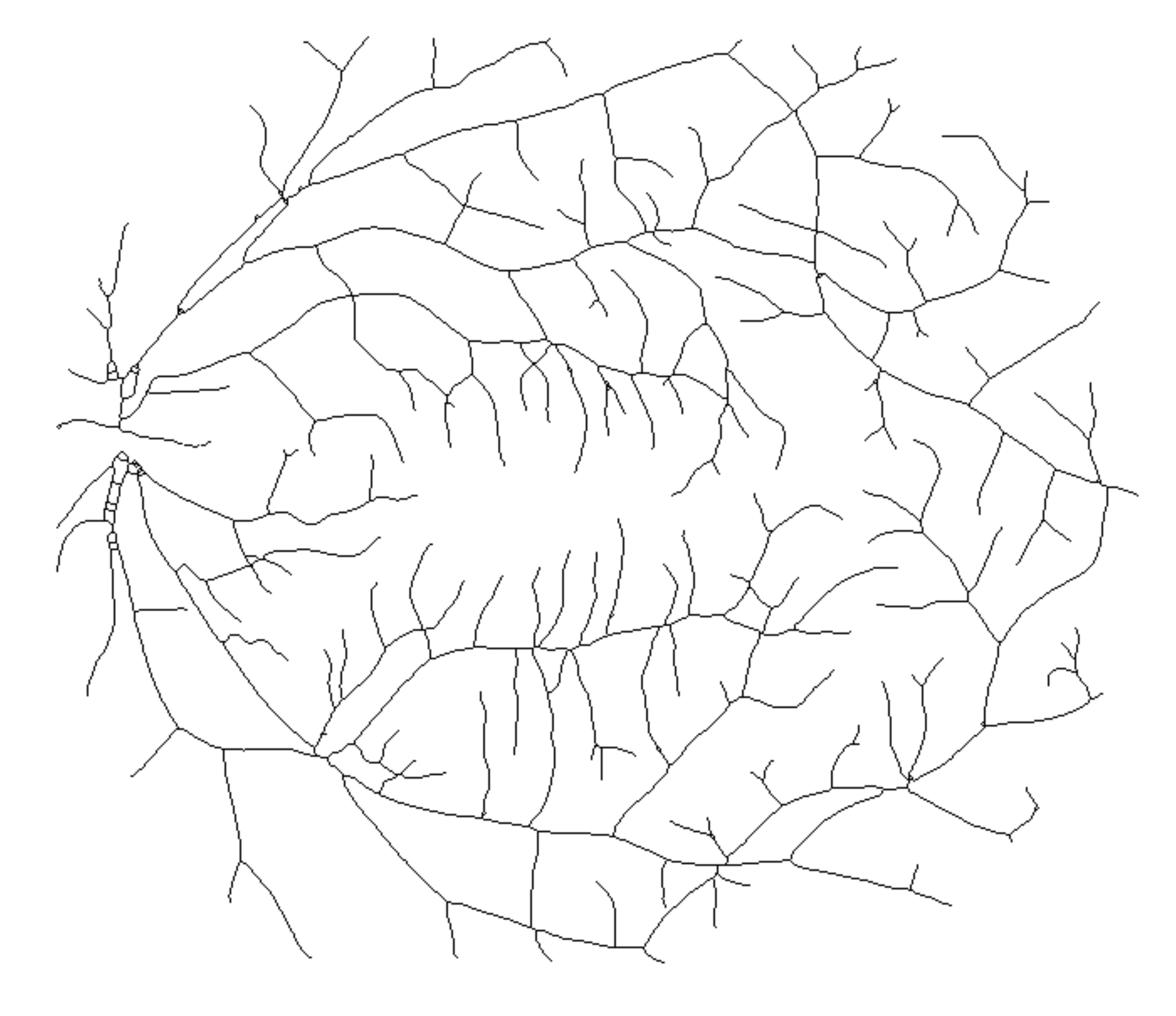}}

	\subfloat[]{\includegraphics[width=0.35\textwidth]{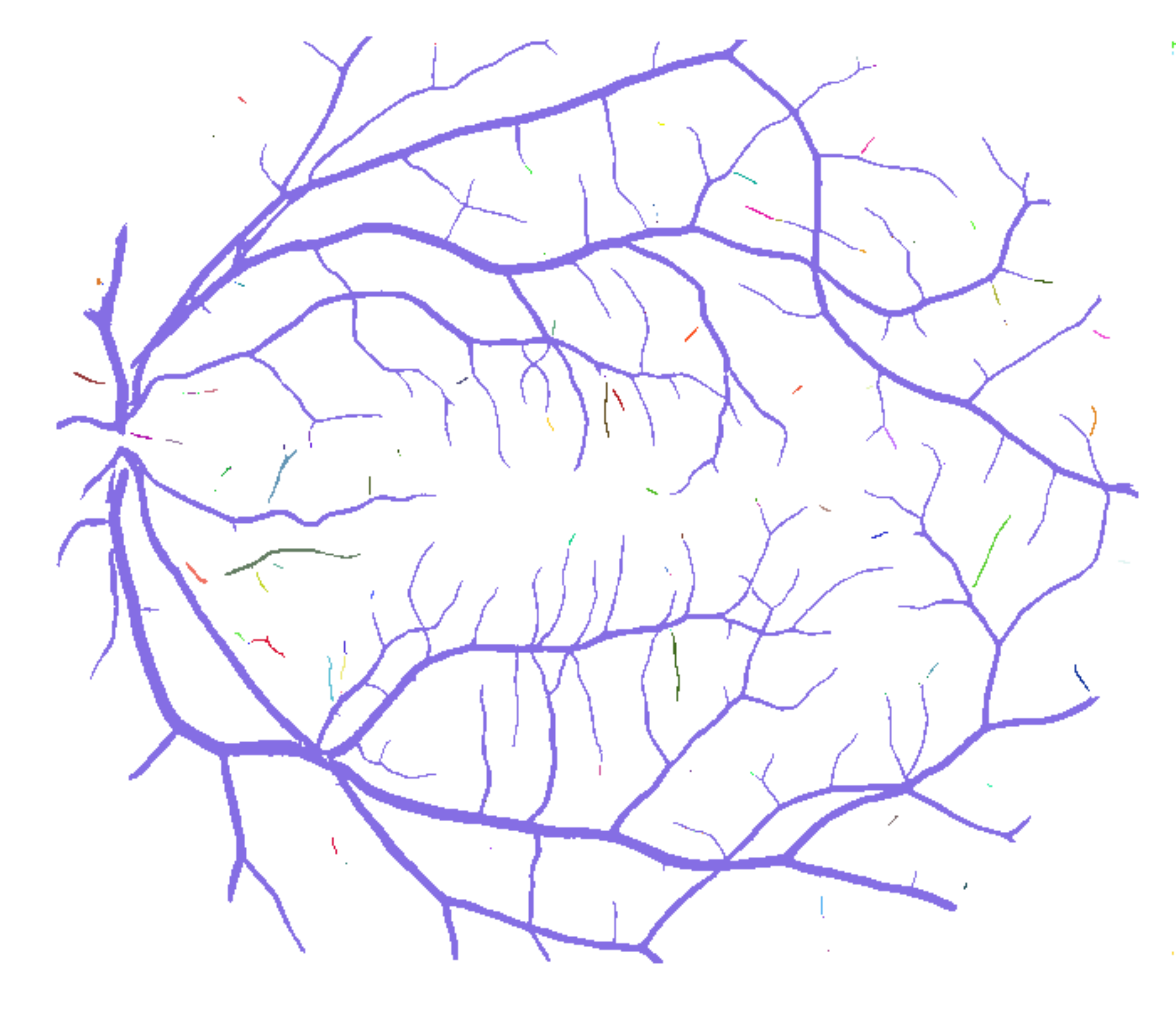}}
	\hfil
	\subfloat[]{\includegraphics[width=0.35\textwidth]{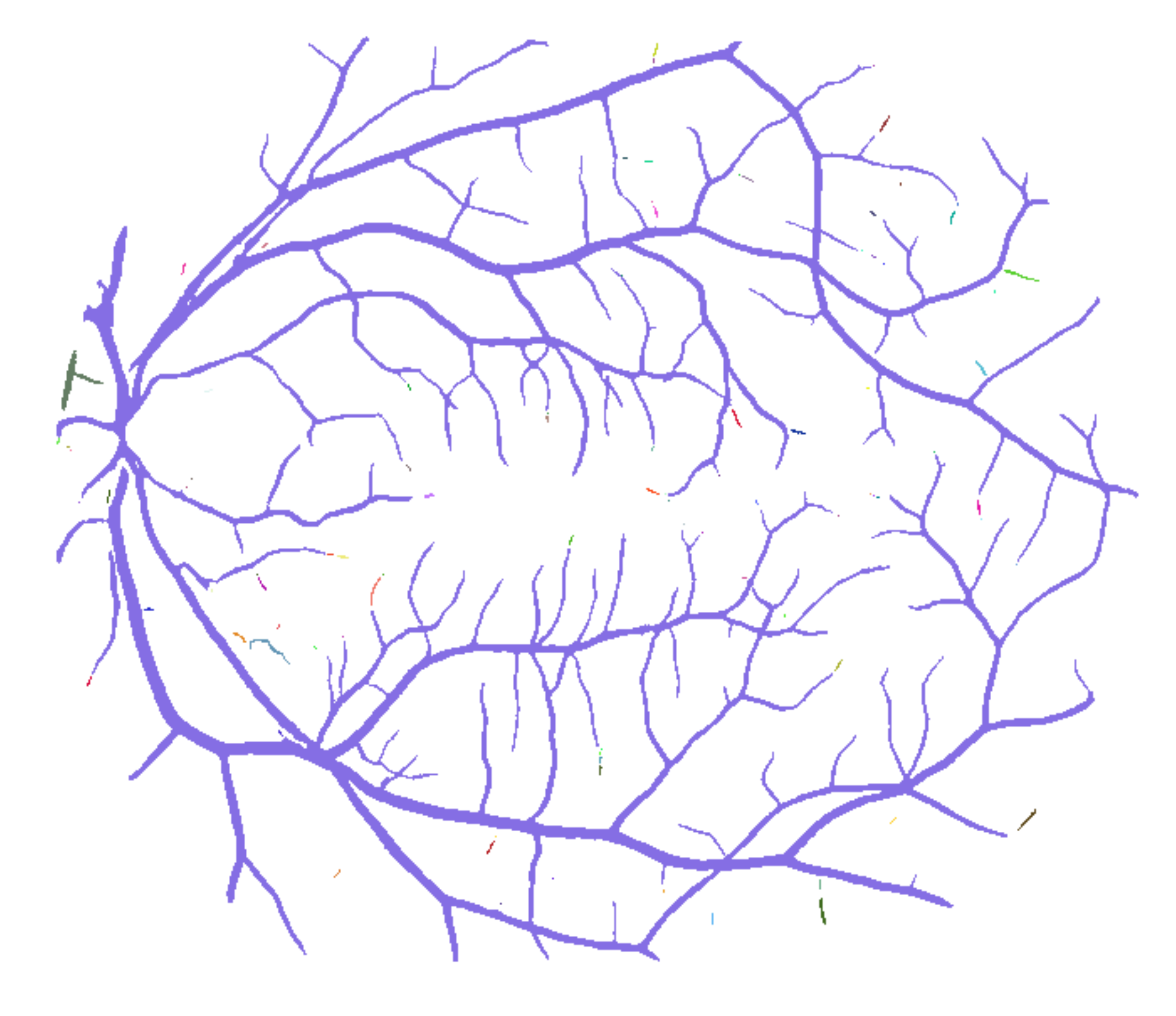}}

	\subfloat[]{\includegraphics[width=0.35\textwidth]{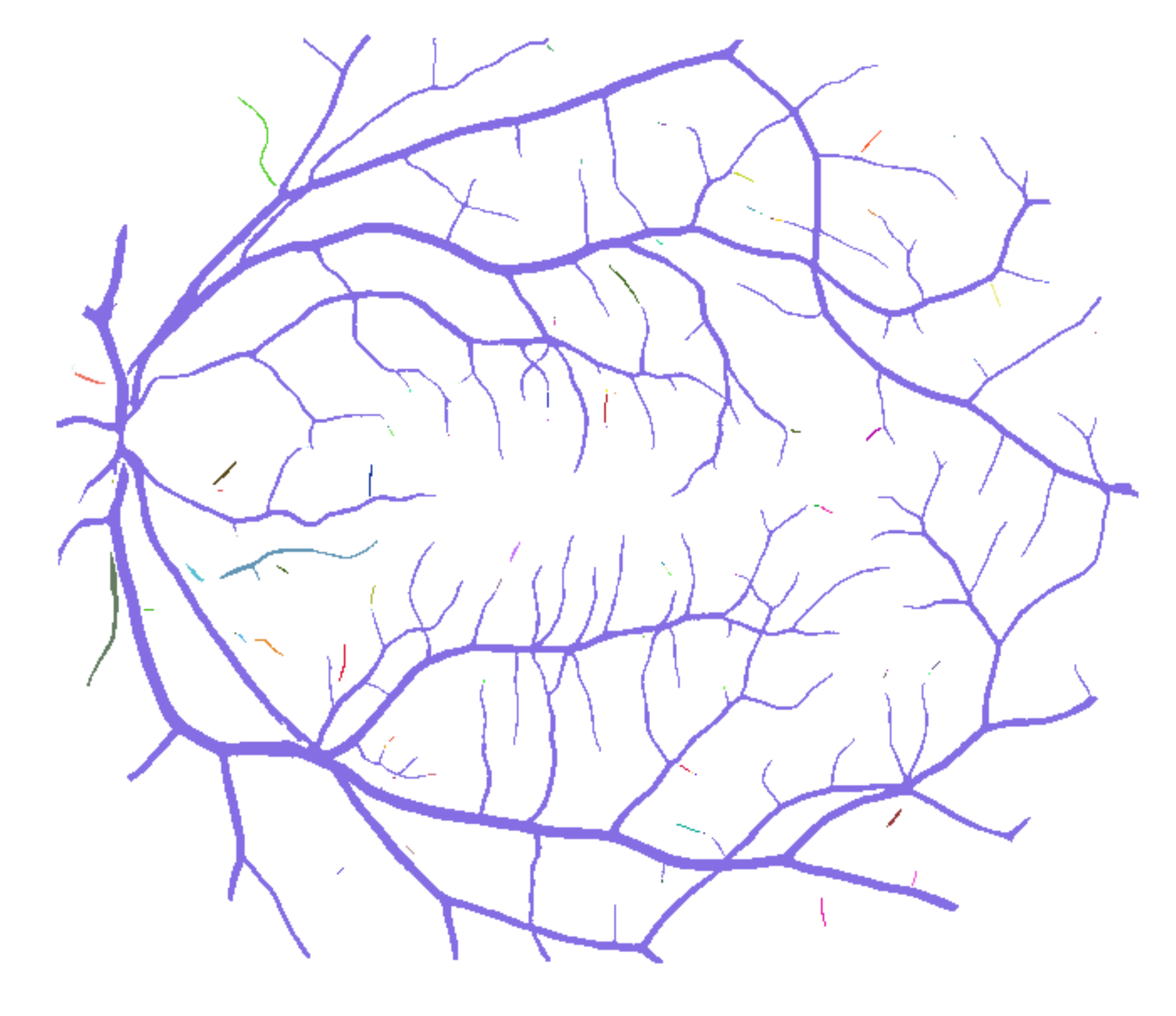}}
	\hfil
	\subfloat[]{\includegraphics[width=0.35\textwidth]{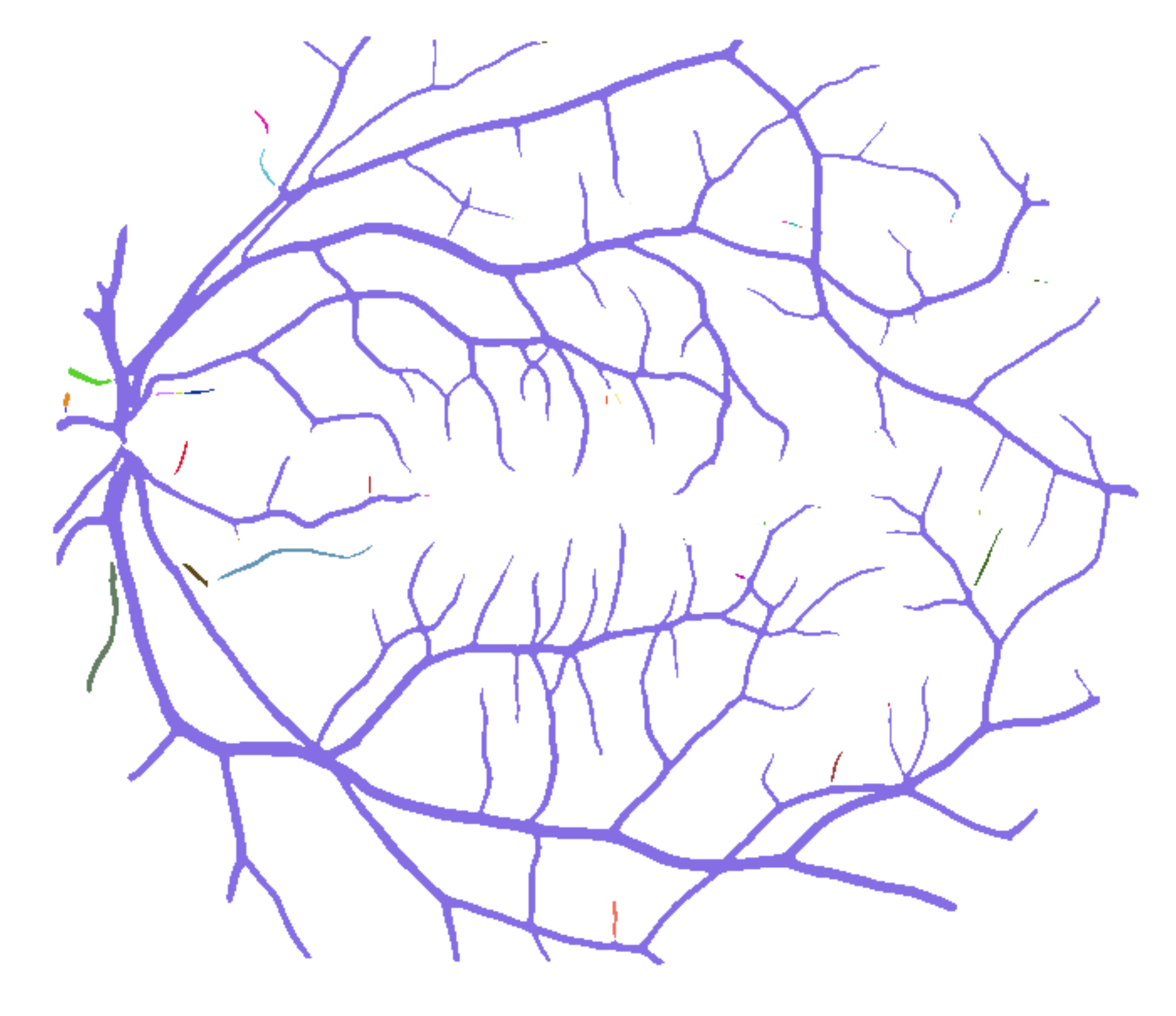}}

	\caption{Vessel segments visualization of a retina image from STARE (when $\text{threshold}=110$ and the connectivity values are provided for each method in the parentheses). (a) Raw image. (b) Extracted center-line from the ground-truth. (c) UNet (0.7128). (d) DenseNet (0.7260). (e) DUNet (0.7095). (f) IterNet (0.9035). Different colors mean different segments. Again, IterNet is the best in connectivity.}
	\label{fig_stare}
\end{figure*}

Figures \ref{fig_drive}, \ref{fig_chasedb1}, and \ref{fig_stare} present the visualization results of the segments in the prediction results. We can see that IterNet almost consistently produces a smaller number of segments.

\newpage

\begin{table*}
	\caption{Performance comparison on the DRIVE dataset (with mask).}
	\label{table_detect_performance_drive1}
	\centering
	\begin{tabular}{l|ccccccc}
		\hline
		Method  & Conn. & F1 Score & Sensitivity & Specificity & Accuracy & AUC\\ 
		\hline
		
		\textbf{IterNet} & \textbf{0.9193} &\textbf{0.8205} & \textbf{0.7735} & 0.9838 &\textbf{0.9573}& \textbf{0.9816}\\
		\textbf{w/o Skip Connection} & 0.9106 & 0.8160 & 0.7659 &  0.9839& 0.9565& 0.9799\\
		\textbf{Iterated UNets } & 0.8893 & 0.8123 & 0.7575 & \textbf{0.9845} & 0.9559& 0.9794\\
		
		\hline
	\end{tabular}
\end{table*}

\begin{table*}
	\caption{Performance comparison on the CHASEDB1 dataset (with mask).}
	\label{table_detect_performance_chase1}
	\centering
	\begin{tabular}{l|ccccccc}
		\hline
		Method  & Conn. & F1 Score & Sensitivity & Specificity & Accuracy & AUC\\ 
		\hline
		
		\textbf{IterNet} & \textbf{0.9091} &  \textbf{0.8073} &  \textbf{0.7970} & 0.9823 & \textbf{0.9655} & \textbf{0.9851}\\
		\textbf{w/o Skip Connection} & 0.8920 & 0.7647 & 0.7001 & \textbf{0.9870} & 0.9610 & 0.9770\\
		\textbf{Iterated UNets } & 0.8773 & 0.7997 & 0.7670 & 0.9849 & 0.9652 & 0.9845\\
		
		\hline
	\end{tabular}
\end{table*}

\begin{table*}
	\caption{Performance comparison on the STARE dataset (with mask).}
	\label{table_detect_performance_stare1}
	\centering
	\begin{tabular}{l|ccccccc}
		\hline
		Method  & Conn. & F1 Score & Sensitivity & Specificity & Accuracy & AUC\\ 
		\hline
		
		\textbf{IterNet} &\textbf{0.8977}&  \textbf{0.8146} &  \textbf{0.7715} & 0.9886 & \textbf{0.9701}& \textbf{0.9881}\\
		\textbf{w/o Skip Connection} & 0.8967 & 0.7482 & 0.6494 & \textbf{0.9920} & 0.9628& 0.9808\\
		\textbf{Iterated UNets } & \textbf{0.8977} &  0.7641 & 0.6764 & 0.9913 & 0.9645 & 0.9830\\
		
		\hline
	\end{tabular}
\end{table*}

\begin{table*}
	\caption{Time costs for prediction of one image using IterNet with and without cropping.}
	\label{table_time_costs}
	\centering
	\begin{tabular}{l|cccccc|c|c}
		\hline
		Method  & Read  & Load Model & Crop & Pred (Patches) & Combine & Write & SUM & AUC\\ 
		\hline
		w. Image Patch (Stride 3) & 8.55s & 2.51s & 2.94s & 58.45s (22801) & 1.03s &  0.01s & 73.49s &\textbf{0.9816}\\
		w. Image Patch (Stride 8) & 8.56s & 2.50s & 0.43s & 10.49s (3249) & 0.16s &  0.01s & 22.15s &0.9815\\
		w. Whole Image. Crop & 8.56s & 2.50s &- & 0.01 (1) & - & 0.01s & \textbf{11.08s} &0.9813\\
		\hline

		\hline
	\end{tabular}
\end{table*}

{\small

}


\begin{thebibliography}{10}
	\providecommand{\url}[1]{#1}
	\csname url@samestyle\endcsname
	\providecommand{\newblock}{\relax}
	\providecommand{\bibinfo}[2]{#2}
	\providecommand{\BIBentrySTDinterwordspacing}{\spaceskip=0pt\relax}
	\providecommand{\BIBentryALTinterwordstretchfactor}{4}
	\providecommand{\BIBentryALTinterwordspacing}{\spaceskip=\fontdimen2\font plus
		\BIBentryALTinterwordstretchfactor\fontdimen3\font minus
		\fontdimen4\font\relax}
	\providecommand{\BIBforeignlanguage}[2]{{%
			\expandafter\ifx\csname l@#1\endcsname\relax
			\typeout{** WARNING: IEEEtran.bst: No hyphenation pattern has been}%
			\typeout{** loaded for the language `#1'. Using the pattern for}%
			\typeout{** the default language instead.}%
			\else
			\language=\csname l@#1\endcsname
			\fi
			#2}}
	\providecommand{\BIBdecl}{\relax}
	\BIBdecl
	
	\bibitem{UNet}
	O.~Ronneberger, P.~Fischer, and T.~Brox, ``{U-Net}: Convolutional networks for
	biomedical image segmentation,'' in \emph{MICCAI}, 2015, pp. 234--241.
	
	\bibitem{chatziralli2012value}
	I.~P. Chatziralli, E.~D. Kanonidou, P.~Keryttopoulos, P.~Dimitriadis, and L.~E.
	Papazisis, ``The value of fundoscopy in general practice,'' \emph{The Open
		Ophthalmology Journal}, vol.~6, pp. 4--5, 2012.
	
	\bibitem{staal:2004-855}
	J.~Staal, M.~Abramoff, M.~Niemeijer, M.~Viergever, and B.~van Ginneken,
	``{Ridge based vessel segmentation in color images of the retina},''
	\emph{{IEEE Transactions on Medical Imaging}}, vol.~23, no.~4, pp. 501--509,
	2004.
	
	\bibitem{owen2009measuring}
	C.~G. Owen, A.~R. Rudnicka, R.~Mullen, S.~A. Barman, D.~Monekosso, P.~H.
	Whincup, J.~Ng, and C.~Paterson, ``Measuring retinal vessel tortuosity in
	10-year-old children: validation of the computer-assisted image analysis of
	the retina ({CAIAR}) program,'' \emph{Investigative Ophthalmology \& Visual
		Science}, vol.~50, no.~5, pp. 2004--2010, 2009.
	
	\bibitem{845178}
	A.~D. {Hoover}, V.~{Kouznetsova}, and M.~{Goldbaum}, ``Locating blood vessels
	in retinal images by piecewise threshold probing of a matched filter
	response,'' \emph{IEEE Transactions on Medical Imaging}, vol.~19, no.~3, pp.
	203--210, 2000.
	
	\bibitem{JIN2019}
	Q.~Jin, Z.~Meng, T.~D. Pham, Q.~Chen, L.~Wei, and R.~Su, ``{DUNet}: A
	deformable network for retinal vessel segmentation,'' \emph{Knowledge-Based
		Systems}, vol. 178, pp. 149--162, 2019.
	
	\bibitem{7319356}
	F.~{Calivá}, M.~{Aletti}, B.~{Al-Diri}, and A.~{Hunter}, ``A new tool to
	connect blood vessels in fundus retinal images,'' in \emph{2015 37th Annual
		International Conference of the IEEE Engineering in Medicine and Biology
		Society (EMBC)}, 2015, pp. 4343--4346.
	
	\bibitem{DOLZ2018456}
	J.~Dolz, C.~Desrosiers, and I.~B. Ayed, ``{3D} fully convolutional networks for
	subcortical segmentation in {MRI}: A large-scale study,'' \emph{NeuroImage},
	vol. 170, pp. 456--470, 2018.
	
	\bibitem{Zhang_2018_CVPR}
	Y.~Zhang, Z.~Qiu, T.~Yao, D.~Liu, and T.~Mei, ``Fully convolutional adaptation
	networks for semantic segmentation,'' in \emph{CVPR}, 2018, pp. 6810--6818.
	
	\bibitem{Jegou_2017_CVPR_Workshops}
	S.~Jegou, M.~Drozdzal, D.~Vazquez, A.~Romero, and Y.~Bengio, ``The one hundred
	layers tiramisu: Fully convolutional densenets for semantic segmentation,''
	in \emph{CVPR Workshops}, 2017, pp. 11--19.
	
	\bibitem{Long_2015_CVPR}
	J.~Long, E.~Shelhamer, and T.~Darrell, ``Fully convolutional networks for
	semantic segmentation,'' in \emph{CVPR}, 2015, pp. 3431--3440.
	
	\bibitem{Amrehn:2017:UIA:3309883.3309905}
	M.~Amrehn, S.~Gaube, M.~Unberath, F.~Schebesch, T.~Horz, M.~Strumia, S.~Steidl,
	M.~Kowarschik, and A.~Maier, ``{UI-Net}: Interactive artificial neural
	networks for iterative image segmentation based on a user model,'' in
	\emph{Proceedings of the Eurographics Workshop on Visual Computing for
		Biology and Medicine (VCBM)}, 2017, pp. 143--147.
	
	\bibitem{DROZDZAL20181}
	M.~Drozdzal, G.~Chartrand, E.~Vorontsov, M.~Shakeri, L.~D. Jorio, A.~Tang,
	A.~Romero, Y.~Bengio, C.~Pal, and S.~Kadoury, ``Learning normalized inputs
	for iterative estimation in medical image segmentation,'' \emph{Medical Image
		Analysis}, vol.~44, pp. 1--13, 2018.
	
	\bibitem{7042289}
	S.~{Roychowdhury}, D.~D. {Koozekanani}, and K.~K. {Parhi}, ``Iterative vessel
	segmentation of fundus images,'' \emph{IEEE Transactions on Biomedical
		Engineering}, vol.~62, no.~7, pp. 1738--1749, 2015.
	
	\bibitem{8341481}
	Z.~{Yan}, X.~{Yang}, and K.~{Cheng}, ``Joint segment-level and pixel-wise
	losses for deep learning based retinal vessel segmentation,'' \emph{IEEE
		Transactions on Biomedical Engineering}, vol.~65, no.~9, pp. 1912--1923,
	2018.
	
	\bibitem{8681706}
	Y.~{Weng}, T.~{Zhou}, Y.~{Li}, and X.~{Qiu}, ``{NAS-Unet}: Neural architecture
	search for medical image segmentation,'' \emph{IEEE Access}, vol.~7, pp.
	44\,247--44\,257, 2019.
	
	\bibitem{8589312}
	X.~{Xiao}, S.~{Lian}, Z.~{Luo}, and S.~{Li}, ``Weighted {Res-UNet} for
	high-quality retina vessel segmentation,'' in \emph{International Conference
		on Information Technology in Medicine and Education (ITME)}, 2018, pp.
	327--331.
	
	\bibitem{DBLP:journals/corr/abs-1903-00923}
	J.~Li, X.~Lin, H.~Che, H.~Li, and X.~Qian, ``Probability map guided
	bi-directional recurrent {UNet} for pancreas segmentation,'' \emph{CoRR},
	vol. abs/1903.00923, 2019.
	
	\bibitem{TANG2019289}
	P.~Tang, Q.~Liang, X.~Yan, S.~Xiang, W.~Sun, D.~Zhang, and G.~Coppola,
	``Efficient skin lesion segmentation using {separable-Unet} with stochastic
	weight averaging,'' \emph{Computer Methods and Programs in Biomedicine}, vol.
	178, pp. 289--301, 2019.
	
	\bibitem{8036917}
	J.~U. {Kim}, H.~G. {Kim}, and Y.~M. {Ro}, ``Iterative deep convolutional
	encoder-decoder network for medical image segmentation,'' in \emph{EMBC},
	2017, pp. 685--688.
	
	\bibitem{8697107}
	S.~{Guan}, A.~{Khan}, S.~{Sikdar}, and P.~{Chitnis}, ``Fully dense {UNet} for
	{2D} sparse photoacoustic tomography artifact removal,'' \emph{IEEE Journal
		of Biomedical and Health Informatics}, 2019 (Early Access).
	
	\bibitem{8379359}
	X.~{Li}, H.~{Chen}, X.~{Qi}, Q.~{Dou}, C.~{Fu}, and P.~{Heng}, ``{H-DenseUNet}:
	Hybrid densely connected {UNet} for liver and tumor segmentation from {CT}
	volumes,'' \emph{IEEE Transactions on Medical Imaging}, vol.~37, no.~12, pp.
	2663--2674, 2018.
	
	\bibitem{8099726}
	G.~{Huang}, Z.~{Liu}, L.~v.~d. {Maaten}, and K.~Q. {Weinberger}, ``Densely
	connected convolutional networks,'' in \emph{CVPR}, 2017, pp. 2261--2269.
	
	\bibitem{dai17dcn}
	J.~Dai, H.~Qi, Y.~Xiong, Y.~Li, G.~Zhang, H.~Hu, and Y.~Wei, ``Deformable
	convolutional networks,'' in \emph{ICCV}, 2017, pp. 764--773.
	
	\bibitem{alom2018recurrent}
	M.~Z. Alom, M.~Hasan, C.~Yakopcic, T.~M. Taha, and V.~K. Asari, ``Recurrent
	residual convolutional neural network based on {U-Net} ({R2U-Net}) for
	medical image segmentation,'' \emph{arXiv preprint arXiv:1802.06955}, 2018.
	
	\bibitem{MOCCIA201871}
	S.~Moccia, E.~D. Momi, S.~E. Hadji, and L.~S. Mattos, ``Blood vessel
	segmentation algorithms — review of methods, datasets and evaluation
	metrics,'' \emph{Computer Methods and Programs in Biomedicine}, vol. 158, pp.
	71--91, 2018.
	
	\bibitem{6019055}
	M.~E. {Gegundez-Arias}, A.~{Aquino}, J.~M. {Bravo}, and D.~{Marin}, ``A
	function for quality evaluation of retinal vessel segmentations,'' \emph{IEEE
		Transactions on Medical Imaging}, vol.~31, no.~2, pp. 231--239, 2012.
	
	\bibitem{kawasaki}
	T.~L. Torp, R.~Kawasaki, T.~Y. Wong, T.~Peto, and J.~Grauslund, ``Temporal
	changes in retinal vascular parameters associated with successful panretinal
	photocoagulation in proliferative diabetic retinopathy: A prospective
	clinical interventional study,'' \emph{Acta Ophthalmologica}, vol.~96, no.~4,
	pp. 405--410, 2018.
	
\end{thebibliography}

\begin{thebibliography}{1}
		\providecommand{\url}[1]{#1}
		\csname url@samestyle\endcsname
		\providecommand{\newblock}{\relax}
		\providecommand{\bibinfo}[2]{#2}
		\providecommand{\BIBentrySTDinterwordspacing}{\spaceskip=0pt\relax}
		\providecommand{\BIBentryALTinterwordstretchfactor}{4}
		\providecommand{\BIBentryALTinterwordspacing}{\spaceskip=\fontdimen2\font plus
			\BIBentryALTinterwordstretchfactor\fontdimen3\font minus
			\fontdimen4\font\relax}
		\providecommand{\BIBforeignlanguage}[2]{{%
				\expandafter\ifx\csname l@#1\endcsname\relax
				\typeout{** WARNING: IEEEtran.bst: No hyphenation pattern has been}%
				\typeout{** loaded for the language `#1'. Using the pattern for}%
				\typeout{** the default language instead.}%
				\else
				\language=\csname l@#1\endcsname
				\fi
				#2}}
		\providecommand{\BIBdecl}{\relax}
		\BIBdecl
		
		\bibitem{Sstaal:2004-855}
		J.~Staal, M.~Abramoff, M.~Niemeijer, M.~Viergever, and B.~van Ginneken,
		``{Ridge based vessel segmentation in color images of the retina},''
		\emph{{IEEE Transactions on Medical Imaging}}, vol.~23, no.~4, pp. 501--509,
		2004.
		
		\bibitem{Sowen2009measuring}
		C.~G. Owen, A.~R. Rudnicka, R.~Mullen, S.~A. Barman, D.~Monekosso, P.~H.
		Whincup, J.~Ng, and C.~Paterson, ``Measuring retinal vessel tortuosity in
		10-year-old children: validation of the computer-assisted image analysis of
		the retina ({CAIAR}) program,'' \emph{Investigative Ophthalmology \& Visual
			Science}, vol.~50, no.~5, pp. 2004--2010, 2009.
		
		\bibitem{S845178}
		A.~D. {Hoover}, V.~{Kouznetsova}, and M.~{Goldbaum}, ``Locating blood vessels
		in retinal images by piecewise threshold probing of a matched filter
		response,'' \emph{IEEE Transactions on Medical Imaging}, vol.~19, no.~3, pp.
		203--210, 2000.
		
	\end{thebibliography}
\end{document}